\pdfoutput=1
\documentclass[10pt,conference]{IEEEtran}
\title{PRISM: A Logical--Physical Boundary Compiler\\for Dual-Rail Qubits}

\usepackage{orcidlink}
\author{\IEEEauthorblockN{%
        Evan McKinney\,\orcidlink{0000-0002-4865-5458}\IEEEauthorrefmark{1}\IEEEauthorrefmark{2}\IEEEauthorrefmark{3},
        Kun Liu\,\orcidlink{0009-0002-9038-8733}\IEEEauthorrefmark{1}\IEEEauthorrefmark{2},
        Harshvardhan K. Babla\,\orcidlink{0009-0009-0601-3500}\IEEEauthorrefmark{2}\IEEEauthorrefmark{3},
        Robert J. Schoelkopf\,\orcidlink{0000-0002-2469-5068}\IEEEauthorrefmark{2}\IEEEauthorrefmark{3},
        Yongshan Ding\,\orcidlink{0000-0002-2338-1315}\IEEEauthorrefmark{1}\IEEEauthorrefmark{2}\IEEEauthorrefmark{3}%
    }
    \IEEEauthorblockA{%
        \IEEEauthorrefmark{1}Department of Computer Science, Yale University, New Haven, CT, USA \\
        \IEEEauthorrefmark{2}Yale Quantum Institute, Yale University, New Haven, CT, USA \\
        \IEEEauthorrefmark{3}Department of Applied Physics, Yale University, New Haven, CT, USA \\
        evan.mckinney@yale.edu, yongshan.ding@yale.edu%
    }
}

\usepackage{graphicx}
\usepackage{amsmath}
\usepackage{amssymb}
\usepackage{amsthm}
\usepackage{mathtools}
\usepackage{bm}

\usepackage{algorithm}        %
\usepackage{algpseudocode}    %

\usepackage{graphicx}
\usepackage{subcaption}
\usepackage{float}
\usepackage{afterpage}
\usepackage{placeins}
\usepackage{booktabs}
\newcommand{\devicerules}{\cmidrule(lr){2-3}\cmidrule(l){4-5}}
\newcommand{\workloadrule}{\cmidrule(r){1-2}}
\usepackage{xcolor}
\usepackage{microtype}

\usepackage[sort, compress]{cite}
\usepackage{hyperref}
\hypersetup{hidelinks}
\makeatletter
\def\url@samestyle{\def\UrlFont{\fontencoding{T1}\selectfont}}
\makeatother

\usepackage{tikz}
\usetikzlibrary{quantikz2}
\usetikzlibrary{positioning,arrows.meta,calc,fit,decorations.pathreplacing,backgrounds,shapes.geometric}

\usepackage{fancyhdr}
\usepackage{silence}
\newcommand{\sysname}{\textsc{Prism}}

\usepackage{xcolor}
\usepackage{ifthen}
\newboolean{showcomments}
\setboolean{showcomments}{true}
\ifthenelse{\boolean{showcomments}}
{ \newcommand{\mynote}[3]{
     \fbox{\bfseries\sffamily\scriptsize#1}
        {\small$\blacktriangleright$\textsf{\emph{\color{#3}{#2}}}$\blacktriangleleft$}}}
{ \newcommand{\mynote}[3]{}}

\definecolor{placeholderpurple}{RGB}{145,45,145}

\hypersetup{
    pdftitle={PRISM: A Logical\textendash Physical Boundary Compiler for Dual-Rail Qubits},
    pdfauthor={Evan McKinney, Kun Liu, Harshvardhan K. Babla, Robert J. Schoelkopf, Yongshan Ding}
}

\begin{document}
\maketitle
\thispagestyle{plain}
\pagestyle{plain}

\begin{abstract}
Quantum computers can encode each logical qubit across multiple physical degrees of freedom so that errors on the physical qubits can be detected or corrected. A compiler must then translate each logical gate into native instructions, and one gate can have several equivalent realizations that interact through different wires. Routing moves those wires, so a realization chosen before routing can force SWAPs that another would avoid, making code expansion and physical routing a phase-ordering problem. \sysname{} therefore keeps each gate's realizations open across the logical--physical boundary and commits to one only when routing reaches a layout that can execute it. We formalize which physical circuits implement a gate, which respect the fault model, and which the compiler keeps, and define when a gate can execute under a layout. For dual-rail cavity qubits, logical $\mathrm{CR}_z$ has four loss-admissible realizations, one per rail pairing, and its logical availability graph is exactly the simple quotient of the occupied device graph by the rail pairs. Across 25 workloads with 7 to 25 logical qubits on three device graphs, \sysname{} reduces routing SWAPs by 20.4--30.8\% and physical depth by 8.8--10.1\% relative to a greedy search that fixes each realization before routing. On the grid and hexagonal lattice, the SWAP reductions hold through 32 logical qubits. By shortening routes, \sysname{} also reduces photon loss on device models and therefore needs 44.3\% fewer attempts per retained shot at 16 logical qubits, as a geometric mean over workloads and device graphs, and it lowers decoded logical error in a three-round Steane memory under skip-gate leakage.
\end{abstract}


\section{Introduction}

\begin{figure}[t]
  \centering
  \definecolor{prismaccent}{RGB}{106,61,154}
\definecolor{beamone}{RGB}{215,48,39}
\definecolor{beamtwo}{RGB}{244,162,15}
\definecolor{beamthree}{RGB}{26,152,80}
\definecolor{beamfour}{RGB}{44,114,184}
\resizebox{\columnwidth}{!}{%
  \begin{tikzpicture}[
    font=\footnotesize,
    node distance=5.0mm,               %
    stage/.style={draw, rounded corners=2pt, align=center,
        inner xsep=3pt, inner ysep=2.5pt,
        minimum height=5.5mm, minimum width=60mm, fill=black!2,
        line width=0.6pt},
    rcirc/.style={anchor=north},
    flow/.style={-{Stealth[length=4pt]}, line width=0.8pt},
    ]

    \node[stage] (alg) {Quantum algorithm};
    \node[stage, below=3.0mm of alg] (log) {Logical transpiler};
    \node[below=of log, minimum width=9mm, minimum height=5.5mm, inner sep=0] (prism) {};

    \node (op) at ($(prism.north)+(-2.8,-0.15)$) {\scalebox{0.75}{%
        \begin{quantikz}[row sep={0.4cm,between origins}, column sep=0.18cm]
          & \ctrl{1} & \\
          & \gate{R_z(\theta)} &
        \end{quantikz}}};

    \fill[prismaccent!12] (prism.north) -- (prism.south west) -- (prism.south east) -- cycle;
    \draw[prismaccent, line width=1pt] (prism.north) -- (prism.south west) -- (prism.south east) -- cycle;
    \node[prismaccent, font=\bfseries\scriptsize, anchor=west] at ([xshift=2mm]prism.east) {\sysname};

    \draw[dashed, black!35, line width=0.5pt]
    ([xshift=-31mm]prism.south) -- ([xshift=31mm]prism.south);
    \node[draw=none, fill=white, text=black!70, inner sep=1.5pt,
      font=\scriptsize\itshape] at ([xshift=20mm]prism.south) {logical--physical boundary};

    \draw[flow] (alg) -- (log);
    \draw[flow] (op.north |- log.south) -- (op.north);
    \draw[flow] (op.east) -- ($(prism.north)!0.5!(prism.south west)$);

    \coordinate (rtop) at ($(prism.south)+(0,-5.0mm)$); %
    \node[rcirc] (r1) at ($(rtop)+(-2.55,0)$) {\scalebox{0.75}{%
        {\color{beamone}\begin{quantikz}[row sep={0.4cm,between origins}, column sep=0.18cm]
              & \ctrl{2} & \\
              & & \\
              & \gate{R_z(\theta)} & \\
              & &
            \end{quantikz}}}};
    \node[rcirc] (r2) at ($(rtop)+(-0.85,0)$) {\scalebox{0.75}{%
        {\color{beamtwo}\begin{quantikz}[row sep={0.4cm,between origins}, column sep=0.18cm]
              & \ctrl{3} & \\
              & & \\
              & & \\
              & \gate{R_z(\text{-}\theta)} &
            \end{quantikz}}}};
    \node[rcirc] (r3) at ($(rtop)+(0.85,0)$) {\scalebox{0.75}{%
        {\color{beamthree}\begin{quantikz}[row sep={0.4cm,between origins}, column sep=0.18cm]
              & & \\
              & \octrl{1} & \\
              & \gate{R_z(\theta)} & \\
              & &
            \end{quantikz}}}};
    \node[rcirc] (r4) at ($(rtop)+(2.55,0)$) {\scalebox{0.75}{%
        {\color{beamfour}\begin{quantikz}[row sep={0.4cm,between origins}, column sep=0.18cm]
              & & \\
              & \octrl{2} & \\
              & & \\
              & \gate{R_z(\text{-}\theta)} &
            \end{quantikz}}}};

    \node[fit=(r1)(r2)(r3)(r4), inner sep=0] (rrow) {};
    \node[stage, below=of rrow] (phys) {Physical transpiler};
    \node[stage, below=3.0mm of phys] (ctrl) {Control pulses};

    \draw[beamone,  line width=1.2pt] (prism.south) -- (r1.north);
    \draw[beamtwo,  line width=1.2pt] (prism.south) -- (r2.north);
    \draw[beamthree,line width=1.2pt] (prism.south) -- (r3.north);
    \draw[beamfour, line width=1.2pt] (prism.south) -- (r4.north);

    \coordinate (feedwire) at ([yshift=4mm]rrow.south);
    \coordinate (feedgate) at ([yshift=2mm]rrow.south);
    \draw[flow, beamone,   line width=1.0pt] (r1.south |- feedwire) -- (r1.south |- phys.north);
    \draw[flow, beamtwo,   line width=1.0pt] (r2.south |- feedgate) -- (r2.south |- phys.north);
    \draw[flow, beamthree, line width=1.4pt] (r3.south |- feedwire) -- (r3.south |- phys.north);
    \draw[flow, beamfour,  line width=1.0pt] (r4.south |- feedgate) -- (r4.south |- phys.north);

    \draw[flow] (phys) -- (ctrl);

  \end{tikzpicture}%
}
  \caption{\sysname{} carries the four realizations of dual-rail
    $\mathrm{CR}_z(\theta)$, one per rail pairing, across the logical--physical boundary for layout-aware
    selection during physical routing. Fig.~\ref{fig:crz-candidates} shows the emitted circuits.}
  \label{fig:overview}
\end{figure}
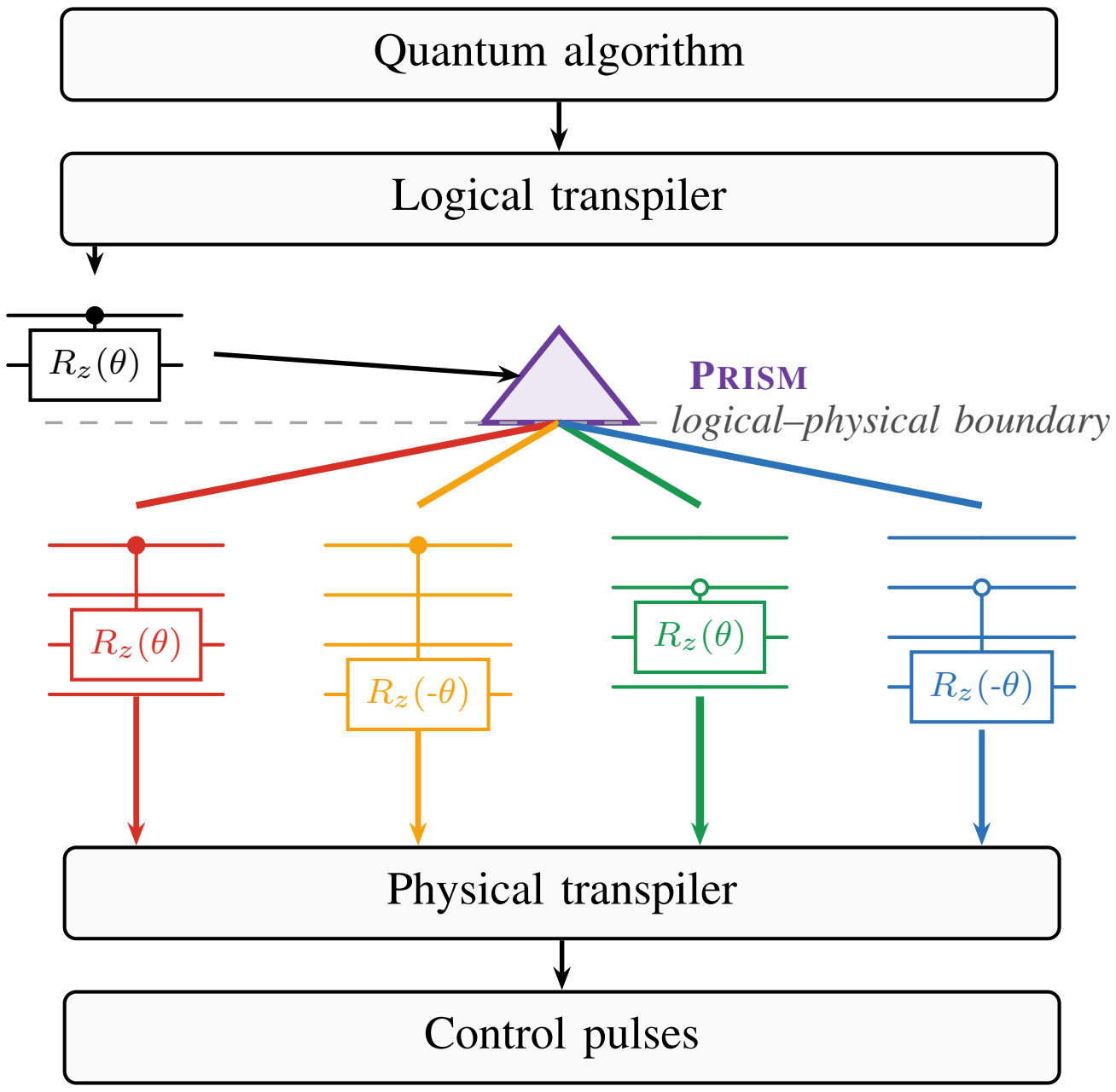

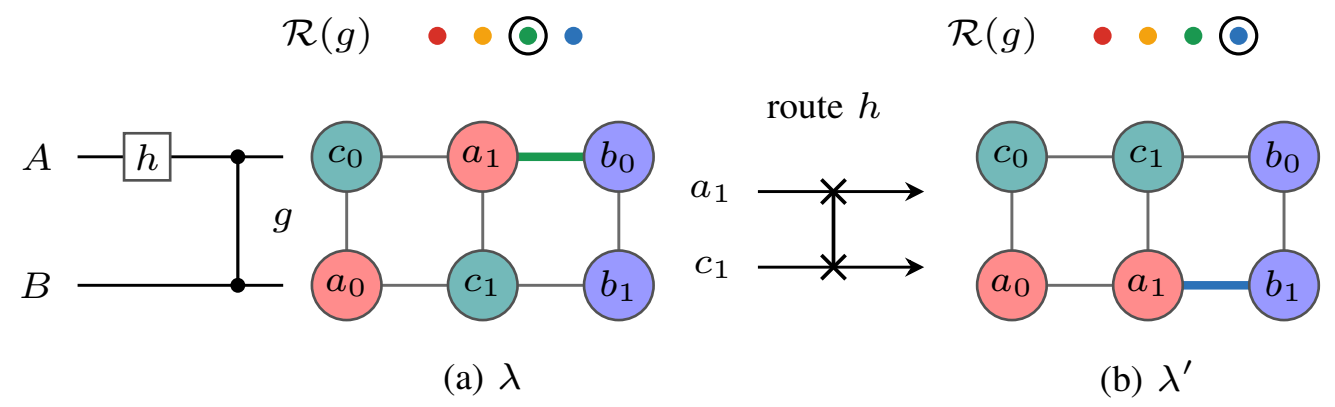
\begin{figure}[t]
  \centering
  \begin{tikzpicture}[
  x=1cm,
  y=1cm,
  font=\footnotesize,
  >={Stealth[length=3.6pt,width=3.6pt]},
  device/.style={line width=.55pt, black!58},
  site/.style={
    circle,
    draw=black!68,
    line width=.5pt,
    minimum size=4.6mm,
    inner sep=0pt,
    font=\scriptsize
  },
]
\definecolor{beamone}{RGB}{215,48,39}
\definecolor{beamtwo}{RGB}{244,162,15}
\definecolor{beamthree}{RGB}{26,152,80}
\definecolor{beamfour}{RGB}{44,114,184}

\node[font=\scriptsize, anchor=east] at (-1.82,0.85) {$A$};
\node[font=\scriptsize, anchor=east] at (-1.82,0) {$B$};
\draw[line width=.55pt] (-1.78,0.85) -- (-0.42,0.85);
\draw[line width=.55pt] (-1.78,0) -- (-0.42,0);
\node[
  draw=black!68,
  fill=white,
  line width=.5pt,
  rectangle,
  inner sep=2pt,
  font=\scriptsize
] at (-1.32,0.85) {$h$};
\draw[line width=.6pt] (-0.72,0.85) -- (-0.72,0);
\fill (-0.72,0.85) circle (1.4pt);
\fill (-0.72,0) circle (1.4pt);
\node[font=\scriptsize, anchor=west] at (-0.62,0.425) {$g$};

\begin{scope}
  \node[anchor=east, font=\scriptsize] at (0.30,1.65) {$\mathcal R(g)$};
  \foreach \cc/\xx in {beamone/0.60, beamtwo/0.90, beamthree/1.20, beamfour/1.50} {
    \fill[\cc] (\xx,1.65) circle (1.7pt);
  }
  \draw[black, line width=.6pt] (1.20,1.65) circle (3.4pt);

  \coordinate (u0) at (0,0.85);
  \coordinate (u1) at (0.9,0.85);
  \coordinate (u2) at (1.8,0.85);
  \coordinate (u3) at (0,0);
  \coordinate (u4) at (0.9,0);
  \coordinate (u5) at (1.8,0);
  \draw[device] (u0)--(u1)--(u2)
                (u3)--(u4)--(u5)
                (u0)--(u3) (u1)--(u4) (u2)--(u5);
  \draw[beamthree, line width=1.6pt] (u1)--(u2);
  \node[site, fill=teal!55] at (u0) {$c_0$};
  \node[site, fill=red!45]  at (u1) {$a_1$};
  \node[site, fill=blue!40] at (u2) {$b_0$};
  \node[site, fill=red!45]  at (u3) {$a_0$};
  \node[site, fill=teal!55] at (u4) {$c_1$};
  \node[site, fill=blue!40] at (u5) {$b_1$};
  \node[font=\scriptsize] at (0.9,-0.62) {(a) $\lambda$};
\end{scope}

\node[font=\scriptsize, anchor=south] at (3.16,0.98) {route $h$};
\node[font=\scriptsize, anchor=east] at (2.68,0.62) {$a_1$};
\node[font=\scriptsize, anchor=east] at (2.68,0.12) {$c_1$};
\draw[->, line width=.55pt] (2.72,0.62) -- (3.82,0.62);
\draw[->, line width=.55pt] (2.72,0.12) -- (3.82,0.12);
\draw[line width=.7pt] (3.22,0.62) -- (3.22,0.12);
\draw[line width=.7pt] (3.14,0.54) -- (3.30,0.70);
\draw[line width=.7pt] (3.14,0.70) -- (3.30,0.54);
\draw[line width=.7pt] (3.14,0.04) -- (3.30,0.20);
\draw[line width=.7pt] (3.14,0.20) -- (3.30,0.04);

\begin{scope}[shift={(4.40,0)}]
  \node[anchor=east, font=\scriptsize] at (0.30,1.65) {$\mathcal R(g)$};
  \foreach \cc/\xx in {beamone/0.60, beamtwo/0.90, beamthree/1.20, beamfour/1.50} {
    \fill[\cc] (\xx,1.65) circle (1.7pt);
  }
  \draw[black, line width=.6pt] (1.50,1.65) circle (3.4pt);

  \coordinate (v0) at (0,0.85);
  \coordinate (v1) at (0.9,0.85);
  \coordinate (v2) at (1.8,0.85);
  \coordinate (v3) at (0,0);
  \coordinate (v4) at (0.9,0);
  \coordinate (v5) at (1.8,0);
  \draw[device] (v0)--(v1)--(v2)
                (v3)--(v4)--(v5)
                (v0)--(v3) (v1)--(v4) (v2)--(v5);
  \draw[beamfour, line width=1.6pt] (v4)--(v5);
  \node[site, fill=teal!55] at (v0) {$c_0$};
  \node[site, fill=teal!55] at (v1) {$c_1$};
  \node[site, fill=blue!40] at (v2) {$b_0$};
  \node[site, fill=red!45]  at (v3) {$a_0$};
  \node[site, fill=red!45]  at (v4) {$a_1$};
  \node[site, fill=blue!40] at (v5) {$b_1$};
  \node[font=\scriptsize] at (0.9,-0.62) {(b) $\lambda'$};
\end{scope}
\end{tikzpicture}%
  \caption{The program applies $h$, a logical $R_X$ on $A$, before the
    pending $\mathrm{CR}_z$ $g$ on $A$ and $B$. Under layout $\lambda$, only the
    green pairing $(a_1,b_0)$ of $g$'s four realizations (dots) is
    available. Routing $h$ makes the rails of $A$ adjacent by exchanging
    $a_1$ and $c_1$, a rail of a third logical qubit $C$, after which only
    the blue pairing $(a_1,b_1)$ is available. A compiler that chooses the
    green realization under $\lambda$ must insert SWAPs to execute it under $\lambda'$.}
  \label{fig:availability-story}
\end{figure}

Choosing a physical circuit for an encoded quantum gate before routing can force SWAPs that an equivalent circuit would avoid. An encoding spreads a logical qubit across several physical degrees of freedom~\cite{preskill1998fault}, and one logical gate can have equivalent physical realizations that interact through different wires. Lowering translates logical operations into physical instructions through a hierarchy of compiler representations~\cite{paz_qstack_2026}, and physical routing moves wires so that each interaction lands on a device edge~\cite{cowtan_routing_2019}. Expanding each gate into one circuit discards the alternatives before routing can evaluate them, and routing first on a fixed logical coupling graph hides the wires that distinguish them. Code expansion and routing therefore form a phase-ordering problem, like those studied among quantum optimization passes~\cite{to_evaluating_2026}.

We introduce \sysname{} (Physical-Realization Instruction-Set Mapping), which keeps each gate's realization set open across the logical--physical boundary and commits to one realization only when the router can execute it. Its boundary intermediate representation (IR) exposes the physical wires while it retains the alternative circuits (Fig.~\ref{fig:overview}). During SABRE-style routing~\cite{li_tackling_2019}, \sysname{} scores the alternatives under the current layout and uses their costs to guide SWAP insertion. Routing changes which realization is available, so commitment should wait for the layout that routing reaches.

A logical qubit occupies the single-excitation subspace $\operatorname{span}\{|01\rangle,|10\rangle\}$ of two cavity modes, or rails, and photon loss leaves a detectable vacuum state~\cite{teoh_dual-rail_2022,chou2024superconducting,koottandavida_erasure_2024}. With local $Z$ corrections, a controlled phase between either rail of one operand and either rail of the other implements the same logical $\mathrm{CR}_z$~\cite{tsunoda_errordetectable_2023,mehta_bias-preserving_2025}. Because the four realizations need different device edges, the layout determines which of them can execute without routing. In Fig.~\ref{fig:availability-story}, only the green pairing of the pending $\mathrm{CR}_z$ is adjacent under the initial layout. The $R_X$ that precedes it needs its operand's rails adjacent, and the SWAP that achieves this breaks the green pairing and creates the blue one. A compiler that has already committed to green must insert SWAPs, whereas \sysname{} selects blue under the layout that routing reaches.

The formalism separates what makes these alternatives interchangeable from what makes them executable. Completions are the physical unitaries with the required action on encoded inputs, admissible completions also satisfy the fault model, and the compiler retains a finite set of realization circuits. A logical gate is available under a layout when one of its realizations can execute without routing. For dual-rail $\mathrm{CR}_z$, the resulting logical availability graph is exactly the simple quotient of the occupied device graph by the rail pairs.

We compare \sysname{} with Greedy, which shares its boundary IR and physical router but fixes each realization at the start of each routing pass, and with Logical-SABRE, which routes on a fixed logical coupling graph before lowering. An ablation measures the effects of choosing a realization under the current layout and of delaying commitment until execution. This paper makes the following contributions.

\begin{itemize}
\item We implement \sysname{}, which carries realization sets through lowering and commits during routing. Across 25 workloads with 7 to 25 logical qubits on three device graphs, it reduces Greedy's routing SWAPs by 20.4--30.8\% and physical depth by 8.8--10.1\%, with SWAP reductions through 32 logical qubits on the grid and hexagonal lattice.
\item We separate logically correct completions, admissible completions, and retained realization sets, and show that the four loss-admissible dual-rail $\mathrm{CR}_z$ realizations induce the simple quotient of the occupied device graph by the rail pairs.
\item On modeled dual-rail cavity hardware, \sysname{} needs 44.3\% fewer attempts per retained shot at 16 logical qubits as a geometric mean over workloads and device graphs, and in a three-round Steane memory under skip-gate leakage it lowers decoded logical error by 13.0--19.6\% with flags at readout and by 22.6--27.7\% with compiled erasure checks.
\end{itemize}

\section{Background}\label{sec:background}

\subsection{Compilation, Routing, and the Phase-Ordering Problem}

General-purpose quantum compilers run synthesis and routing passes on circuits with explicit wires~\cite{javadiabhari_qiskit_2024,sivarajah_tket_2020,younis_bqskit_2021}. Physical routing maps these wires to a device graph and inserts SWAPs for unsupported interactions~\cite{cowtan_routing_2019}. SABRE scores candidate SWAPs by ready-gate distances and a lookahead set~\cite{li_tackling_2019}.

Multi-level compiler frameworks represent changes in instruction and wire granularity across compilation stages~\cite{lattner_mlir_2021,mccaskey_mlir_2021,paz_qstack_2026,qiskit_multistage_2026}. Realization choice and routing constrain each other, creating a phase-ordering problem~\cite{vegdahl_phase_1982}.

\subsection{Dual-Rail Qubits}

A dual-rail qubit uses two physical modes, or rails, with logical states encoded as $|0_{\rm L}\rangle=|01\rangle$ and $|1_{\rm L}\rangle=|10\rangle$. Together these span the two-level code space $\mathcal{C}=\mathrm{span}\{|01\rangle,|10\rangle\}$, and its complement $\mathcal{C}^{\perp}=\mathrm{span}\{|00\rangle,|11\rangle\}$ defines the erasure space. The encoding can be summarized by the isometry
\begin{equation}
  V \;=\; |01\rangle\langle 0| \;+\; |10\rangle\langle 1|,
  \qquad V^{\dagger} V = I_2, \quad VV^{\dagger}=P_{\mathcal{C}}.
  \label{eq:dr-isometry}
\end{equation}
Under the zero-temperature amplitude-damping model, photon loss maps both logical basis states to the vacuum $|00\rangle$, and erasure detection has been demonstrated in dual-rail cavities, dual-rail transmons, and neutral atoms~\cite{chou2024superconducting,huang_logical_2026,ma_midcircuit_2023,scholl_erasure_2023}. Mid-circuit erasure checks must measure the joint photon occupation of both rails without revealing the logical state~\cite{teoh_dual-rail_2022,degraaf_midcircuit_2025}. At the end of a circuit, each rail can be measured separately~\cite{chou2024superconducting}. Outcomes $01$ and $10$ report logical values, while $00$ and $11$ flag an erasure.

Postselection discards any shot with a detected loss, so erasure probability sets the shot-acceptance rate~\cite{griffiths_limits_2026}. An outer quantum error correction (QEC) code can instead use erasure information to correct the loss~\cite{wu_erasure_2022,kubica_erasure_2022,violaris_erasure_2026,chang_imperfect_2025}, since a code corrects twice as many erasures as Pauli errors~\cite{grassl_erasure_1997} and erasure conversion improves thresholds and overheads~\cite{gu_faulttolerant_2025,gu_optimizing_2025}.

A surface code with erasure-flagging readout can recover erasure-like error scaling without mid-circuit checks~\cite{pavlovich_erasure_2026}. Section~\ref{sec:outer-code} evaluates Steane-code memory circuits with flags at readout and with compiled erasure checks.

The circuit-QED architecture model in this paper uses long-lived cavity modes as rails and an ancilla transmon for joint erasure checks~\cite{teoh_dual-rail_2022,chou2024superconducting,koottandavida_erasure_2024}. Parametric beamsplitters transfer excitations between cavities, including a full cavity--cavity exchange~\cite{chapman_beamsplitter_2023,lu_beamsplitting_2023}. The SWAP--wait--SWAP (SWS) sequence moves a selected control-rail excitation into a transmon coupler, accumulates a phase conditional on one target rail, and returns it~\cite{mehta_bias-preserving_2025}. After local $Z$ corrections, it implements the physical $\mathrm{CR}_z$ instruction of Section~\ref{sec:two-qubit-completions}.

Photon loss dominates errors in dual-rail cavity hardware, occurring during idle periods about six times more often than phase errors and at least 150 times more often than bit flips, and erasure checks identify more than 99\% of losses~\cite{chou2024superconducting}. The SWS gate has approximately $0.5\%$ erasure probability, at most $0.1\%$ residual dephasing, and a bit-flip probability of a few parts per million~\cite{mehta_bias-preserving_2025}, so the residual unflagged errors are mostly phase errors.

The routing model represents each cavity mode as a vertex and each supported two-mode interaction as an edge, and assumes that every edge supports the SWS interaction and a tunable beamsplitter that implements each routing SWAP as a full cavity--cavity exchange.

\section{From Logical Gates to Logical Availability}\label{sec:availability-framework}

Whether an encoded gate can execute without routing depends on its retained circuits and the current layout. For the dual-rail cavity instruction set architecture (ISA), logical availability is a quotient of the occupied device graph by the rail pairs.

\subsection{Completions, Admissibility, and Realizations}\label{sec:lowering-partial-spec}

Let $V$ be the isometry for the encoded operands on which a logical gate $g$ acts, including any fixed ancilla input state. A physical unitary $U$ is a \emph{completion} of $g$ when
\begin{equation}
  UV=Vg.
  \label{eq:physical-completion}
\end{equation}
The corresponding completion set is
\begin{equation}
  \mathcal{P}(g)=\{U:UV=Vg\}.
  \label{eq:completion-set}
\end{equation}
The \emph{admissible completion set} $\mathcal{A}(g)\subseteq\mathcal{P}(g)$ contains completions satisfying the fault model's constraints on the full target-level unitary~\cite{puri_biaspreserving_2020,reinhold_errorcorrected_2020,tsunoda_errordetectable_2023}.

The compiler retains a finite \emph{realization set} $\mathcal{R}(g)$ of circuits, chosen for example by synthesis cost~\cite{mckinney_gulps_2025}. Each realization $r\in\mathcal{R}(g)$ names a complete target-level circuit $C_r$ whose unitary lies in $\mathcal{A}(g)$ and whose implementation obeys the declared circuit-level constraints.

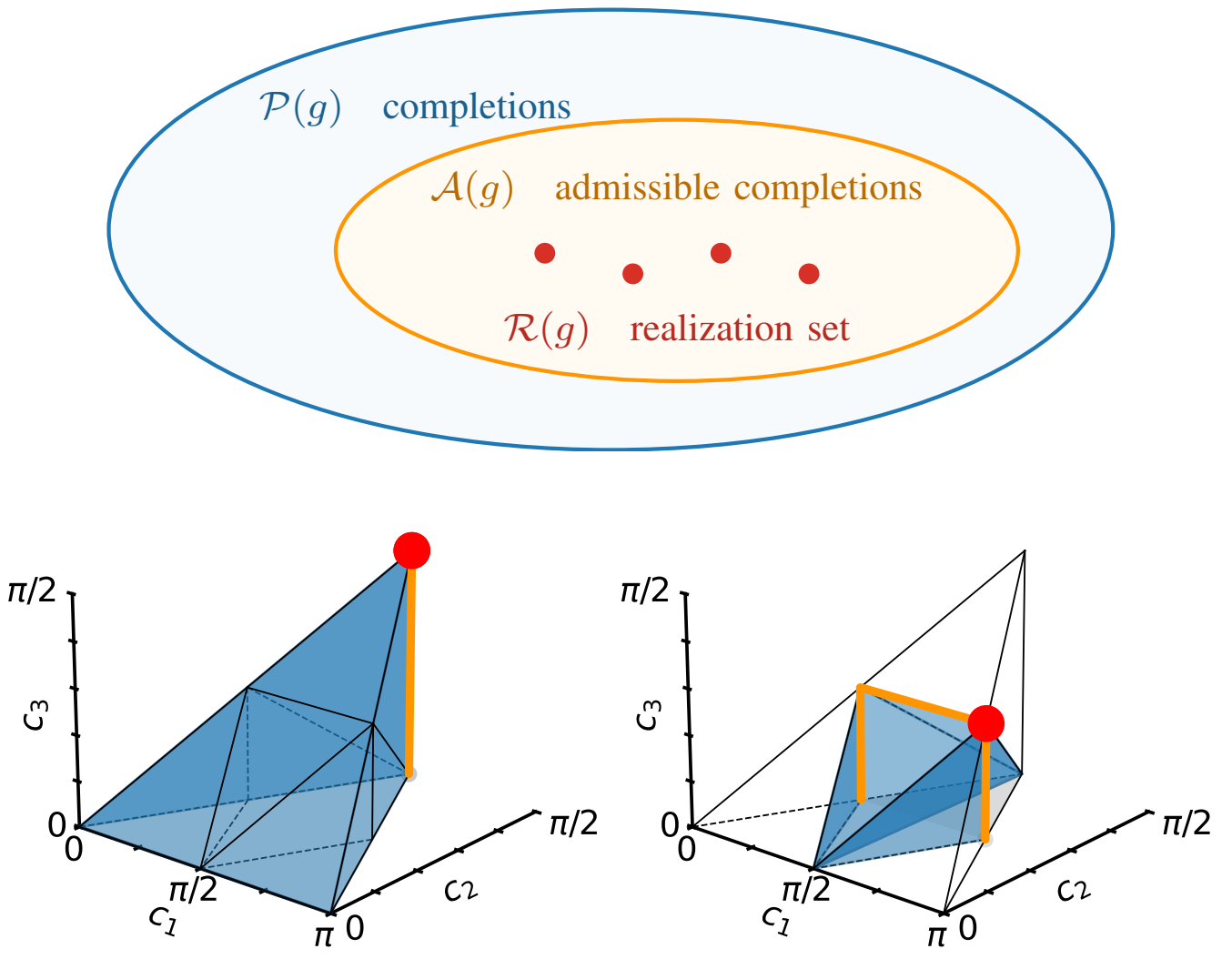
\begin{figure}[t]
  \centering
  \begin{tikzpicture}[font=\footnotesize]
  \definecolor{completionblue}{RGB}{31,119,180}
  \definecolor{admissibleorange}{RGB}{255,149,0}
  \definecolor{realizationred}{RGB}{215,48,39}

  \filldraw[
    fill=completionblue!4,
    draw=completionblue,
    line width=0.8pt
  ] (0,0) ellipse [x radius=3.65cm, y radius=1.60cm];

  \filldraw[
    fill=admissibleorange!5,
    draw=admissibleorange,
    line width=0.8pt
  ] (0.48,-0.15) ellipse [x radius=2.48cm, y radius=0.95cm];

  \node[
    text=completionblue!80!black,
    anchor=west
  ] at (-2.68,0.88)
    {$\mathcal P(g)$\quad completions};

  \node[
    text=admissibleorange!72!black
  ] at (0.48,0.29)
    {$\mathcal A(g)$\quad admissible completions};

  \foreach \x/\y in {-0.48/-0.17,0.16/-0.32,0.80/-0.17,1.44/-0.32}
    \fill[realizationred] (\x,\y) circle (2.15pt);

  \node[
    text=realizationred!85!black
  ] at (0.48,-0.73)
    {$\mathcal R(g)$\quad realization set};
\end{tikzpicture}
  \vspace{3pt}
  \includegraphics[width=\linewidth]{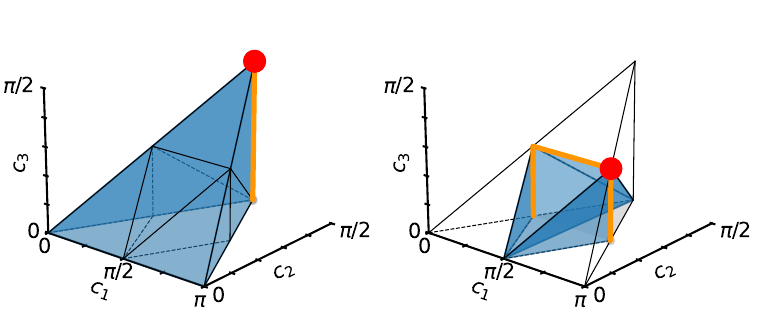}
  \caption{The upper panel distinguishes the physical completions
    $\mathcal{P}(g)$, the admissible completions $\mathcal{A}(g)$, and the
    unitary $U(C_r)$ of each realization $r$ in $\mathcal{R}(g)$, drawn as a
    red point. The lower panels show these sets in the Weyl chamber for
    $U_L=X$ (left) and $U_L=\sqrt{X}$ (right), where the completions form the
    blue surface, the admissible family of
    Eq.~\eqref{eq:loss-compatible-complement} forms the orange line, and the
    realization is the red point.}
  \label{fig:weyl-sweep}
\end{figure}

\subsection{Logical Availability}\label{sec:logical-availability}
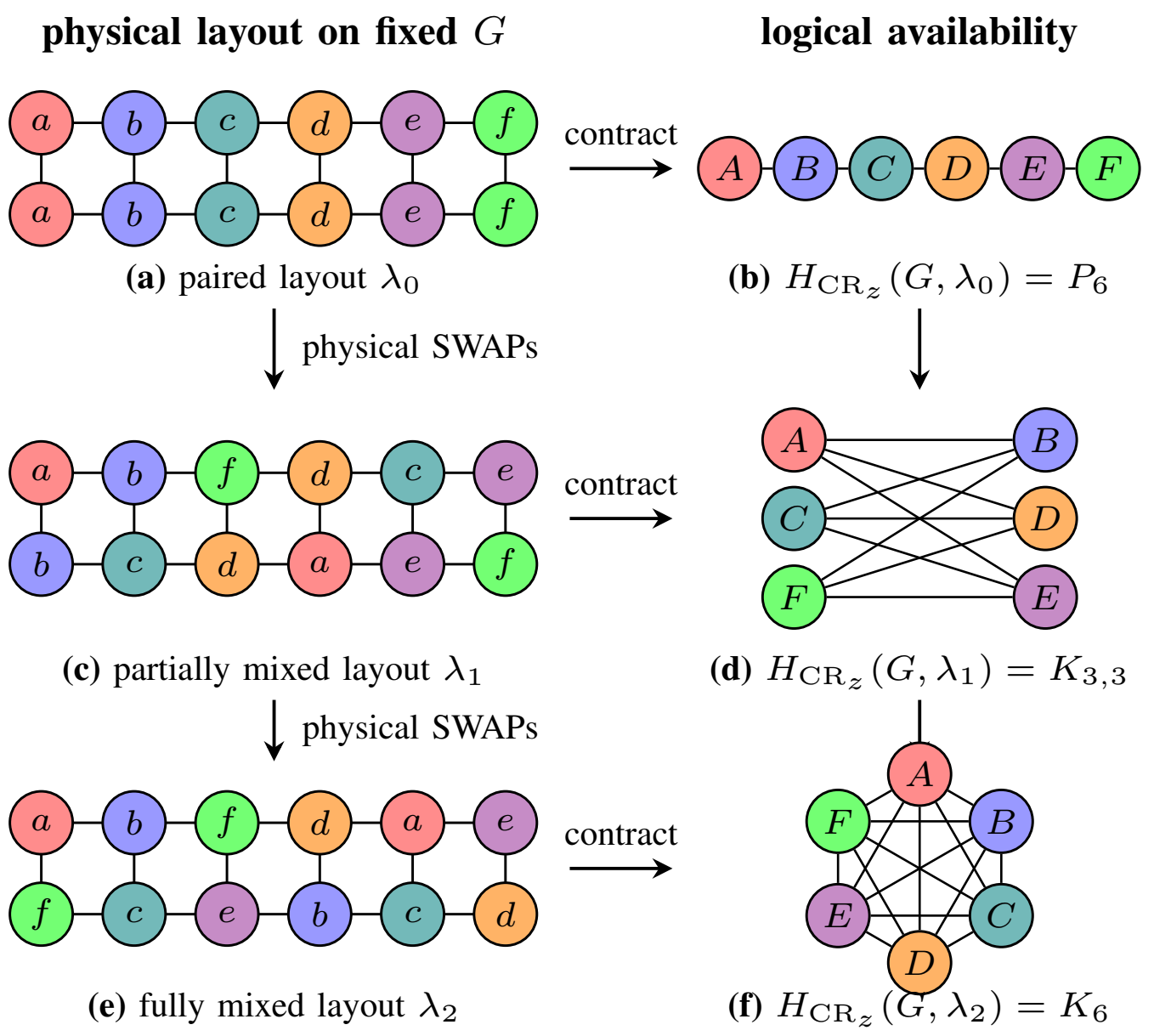
\begin{figure}[t]
  \centering
  \begingroup
\resizebox{\columnwidth}{!}{%
\begin{tikzpicture}[
  font=\footnotesize,
  >={Stealth[length=4pt,width=4pt]},
  site/.style={circle,draw=black,line width=.5pt,minimum size=4.4mm,inner sep=0pt,font=\scriptsize},
  logical/.style={circle,draw=black,line width=.5pt,minimum size=4.4mm,inner sep=0pt,font=\scriptsize},
  device/.style={black,line width=.5pt},
  qedge/.style={black,line width=.45pt},
  flow/.style={->,line width=.65pt},
  change/.style={->,line width=.65pt},
  rowlabel/.style={font=\scriptsize},
]

\node[font=\footnotesize\bfseries] at (1.63,.94)
  {physical layout on fixed $G$};
\node[font=\footnotesize\bfseries] at (6.15,.94)
  {logical availability};

\newcommand{\quotientladder}[3]{%
  \begin{scope}[shift={(0,#1)}]
    \foreach \x in {0,...,5}{
      \coordinate (t\x#2) at (.65*\x,.32);
      \coordinate (b\x#2) at (.65*\x,-.32);
    }
    \draw[device]
      (t0#2)--(t1#2)--(t2#2)--(t3#2)--(t4#2)--(t5#2)
      (b0#2)--(b1#2)--(b2#2)--(b3#2)--(b4#2)--(b5#2);
    \foreach \x in {0,...,5}{\draw[device] (t\x#2)--(b\x#2);}
    #3
  \end{scope}%
}

\quotientladder{0}{i}{
  \node[site,fill=red!45]    at (t0i) {$a$};
  \node[site,fill=blue!40]   at (t1i) {$b$};
  \node[site,fill=teal!55]   at (t2i) {$c$};
  \node[site,fill=orange!60] at (t3i) {$d$};
  \node[site,fill=violet!45] at (t4i) {$e$};
  \node[site,fill=green!55]  at (t5i) {$f$};
  \node[site,fill=red!45]    at (b0i) {$a$};
  \node[site,fill=blue!40]   at (b1i) {$b$};
  \node[site,fill=teal!55]   at (b2i) {$c$};
  \node[site,fill=orange!60] at (b3i) {$d$};
  \node[site,fill=violet!45] at (b4i) {$e$};
  \node[site,fill=green!55]  at (b5i) {$f$};
}
\draw[flow] (3.70,0)--node[above=1pt,font=\scriptsize]{contract}(4.42,0);
\begin{scope}[shift={(4.82,0)}]
  \node[logical,fill=red!45]    (p0) at (0,0) {$A$};
  \node[logical,fill=blue!40]   (p1) at (.53,0) {$B$};
  \node[logical,fill=teal!55]   (p2) at (1.06,0) {$C$};
  \node[logical,fill=orange!60] (p3) at (1.59,0) {$D$};
  \node[logical,fill=violet!45] (p4) at (2.12,0) {$E$};
  \node[logical,fill=green!55]  (p5) at (2.65,0) {$F$};
  \draw[qedge] (p0)--(p1)--(p2)--(p3)--(p4)--(p5);
\end{scope}
\node[rowlabel] at (1.63,-.78)
  {\textbf{(a)} paired layout $\lambda_0$};
\node[rowlabel] at (6.15,-.78)
  {\textbf{(b)} $H_{\mathrm{CR}_z}(G,\lambda_0)=P_6$};

\draw[change] (1.63,-.98)--node[right=2pt,font=\scriptsize]
  {physical SWAPs}(1.63,-1.55);
\draw[change] (6.15,-.98)--(6.15,-1.55);

\quotientladder{-2.45}{m}{
  \node[site,fill=red!45]    at (t0m) {$a$};
  \node[site,fill=blue!40]   at (t1m) {$b$};
  \node[site,fill=green!55]  at (t2m) {$f$};
  \node[site,fill=orange!60] at (t3m) {$d$};
  \node[site,fill=teal!55]   at (t4m) {$c$};
  \node[site,fill=violet!45] at (t5m) {$e$};
  \node[site,fill=blue!40]   at (b0m) {$b$};
  \node[site,fill=teal!55]   at (b1m) {$c$};
  \node[site,fill=orange!60] at (b2m) {$d$};
  \node[site,fill=red!45]    at (b3m) {$a$};
  \node[site,fill=violet!45] at (b4m) {$e$};
  \node[site,fill=green!55]  at (b5m) {$f$};
}
\draw[flow] (3.70,-2.45)--node[above=1pt,font=\scriptsize]
  {contract}(4.42,-2.45);
\begin{scope}[shift={(5.27,-2.45)}]
  \node[logical,fill=red!45]    (k0) at (0,.55) {$A$};
  \node[logical,fill=teal!55]   (k2) at (0,0) {$C$};
  \node[logical,fill=green!55]  (k5) at (0,-.55) {$F$};
  \node[logical,fill=blue!40]   (k1) at (1.76,.55) {$B$};
  \node[logical,fill=orange!60] (k3) at (1.76,0) {$D$};
  \node[logical,fill=violet!45] (k4) at (1.76,-.55) {$E$};
  \foreach \a in {k0,k2,k5}
    \foreach \b in {k1,k3,k4}
      \draw[qedge] (\a)--(\b);
\end{scope}
\node[rowlabel] at (1.63,-3.52)
  {\textbf{(c)} partially mixed layout $\lambda_1$};
\node[rowlabel] at (6.15,-3.52)
  {\textbf{(d)} $H_{\mathrm{CR}_z}(G,\lambda_1)=K_{3,3}$};

\draw[change] (1.63,-3.72)--node[right=2pt,font=\scriptsize]
  {physical SWAPs}(1.63,-4.15);
\draw[change] (6.15,-3.72)--(6.15,-4.15);

\quotientladder{-4.90}{f}{
  \node[site,fill=red!45]    at (t0f) {$a$};
  \node[site,fill=blue!40]   at (t1f) {$b$};
  \node[site,fill=green!55]  at (t2f) {$f$};
  \node[site,fill=orange!60] at (t3f) {$d$};
  \node[site,fill=red!45]    at (t4f) {$a$};
  \node[site,fill=violet!45] at (t5f) {$e$};
  \node[site,fill=green!55]  at (b0f) {$f$};
  \node[site,fill=teal!55]   at (b1f) {$c$};
  \node[site,fill=violet!45] at (b2f) {$e$};
  \node[site,fill=blue!40]   at (b3f) {$b$};
  \node[site,fill=teal!55]   at (b4f) {$c$};
  \node[site,fill=orange!60] at (b5f) {$d$};
}
\draw[flow] (3.70,-4.90)--node[above=1pt,font=\scriptsize]
  {contract}(4.42,-4.90);
\begin{scope}[shift={(6.15,-4.90)}]
  \node[logical,fill=red!45]    (a0) at (90:.66)  {$A$};
  \node[logical,fill=blue!40]   (a1) at (30:.66)  {$B$};
  \node[logical,fill=teal!55]   (a2) at (-30:.66) {$C$};
  \node[logical,fill=orange!60] (a3) at (-90:.66) {$D$};
  \node[logical,fill=violet!45] (a4) at (-150:.66){$E$};
  \node[logical,fill=green!55]  (a5) at (150:.66) {$F$};
  \foreach \a/\b in {
    a0/a1,a0/a2,a0/a3,a0/a4,a0/a5,
    a1/a2,a1/a3,a1/a4,a1/a5,
    a2/a3,a2/a4,a2/a5,
    a3/a4,a3/a5,a4/a5}
    \draw[qedge] (\a)--(\b);
\end{scope}
\node[rowlabel] at (1.63,-5.88)
  {\textbf{(e)} fully mixed layout $\lambda_2$};
\node[rowlabel] at (6.15,-5.88)
  {\textbf{(f)} $H_{\mathrm{CR}_z}(G,\lambda_2)=K_6$};

\end{tikzpicture}%
}
\endgroup
  \caption{Panels (a), (c), and (e) show layouts of six dual-rail qubits
    $A$ to $F$ before and after successive SWAPs, and panels (b), (d), and
    (f) show their logical availability graphs. Each SWAP interleaves rails
    of different rail pairs, and the graph grows from $P_6$ to $K_{3,3}$ to
    $K_6$.}
  \label{fig:availability-graph}
\end{figure}

A logical gate is \emph{available} under a layout when at least one of its realizations can execute without routing. Because routing moves the wires that each realization uses, it changes availability even while the device graph stays fixed (Fig.~\ref{fig:availability-graph}).

Each realization $r$ has an \emph{interaction pattern} $I_r$ with one vertex for each wire of $C_r$ and one labeled edge for each interaction between wires. Let $G$ be the device graph, and let a layout $\lambda$ map the program's physical data wires to vertices in $G$. An embedding $\phi\in\operatorname{Emb}_{\lambda}(I_r,G)$ maps the wires of $C_r$ injectively to device vertices. It fixes each data wire $w$ at $\lambda(w)$ and maps each labeled interaction to a supporting device edge. The \emph{availability relation} under the current layout is
\begin{equation}
  \mathcal{E}_{g}(G,\lambda)
  =
  \{(r,\phi):r\in\mathcal{R}(g),\
  \phi\in\operatorname{Emb}_{\lambda}(I_r,G)\}.
  \label{eq:availability-relation}
\end{equation}
For a two-operand gate type $g$, the \emph{logical availability graph} $H_g(G,\lambda)$ has one vertex for each logical qubit and an edge $(p,q)$ when Eq.~\eqref{eq:availability-relation} is nonempty for $g$ on those operands.

\subsection{Dual-Rail Single-Qubit Completions}

Up to global phase, a logical single-qubit gate has a completion that acts on each rail separately exactly when it is $R_Z(\theta)$ or $R_Z(\theta)X$. Of these gates, only $R_Z(\theta)$ has an admissible completion that acts on each rail separately. In the dual-rail cavity ISA, $R_Z$ therefore executes on a single rail while the two rails of a qubit are separated, whereas $R_X$ requires the two rails to be adjacent, so an upcoming $R_X$ limits how long the router can keep that rail pair separated. We write logical rotations in upper case ($R_X$, $R_Z$) and physical rotations in lower case ($R_z$).

A physical gate that preserves the code and erasure spaces has the block form~\cite{brod2011extending}
\begin{equation}
  M(A, B) = \begin{pmatrix} A_{11} & 0 & 0 & A_{12} \\ 0 & B_{11} & B_{12} & 0 \\ 0 & B_{21} & B_{22} & 0 \\ A_{21} & 0 & 0 & A_{22} \end{pmatrix},
  \label{eq:GAB}
\end{equation}
where $B=U_L$ acts on $\mathcal{C}=\operatorname{span}\{|01\rangle,|10\rangle\}$ and $A$ acts on $\mathcal{C}^{\perp}=\operatorname{span}\{|00\rangle,|11\rangle\}$. Figure~\ref{fig:weyl-sweep} places the two-rail completions of a logical single-qubit gate in the Weyl chamber~\cite{zhang2003geometric,tucci_introduction_2005}, which groups two-qubit unitaries by nonlocal action. The origin represents product unitaries. The blue surface for logical $X$ reaches the origin, so a product unitary can implement $X$. The surface for logical $\sqrt{X}$ does not reach the origin, so every completion of $\sqrt{X}$ entangles the rails. Appendix~\ref{app:weyl-derivation} derives the surfaces.

It follows from the canonical form of parity-preserving gates~\cite{brod2011extending} that $M(A,B)$ is a product unitary only when $A$ and $B$ are both diagonal or both anti-diagonal. Thus, up to global phase, the logical gates with product completions are
\begin{equation}
  \mathcal{G}_{\mathrm{deloc}} = \{R_Z(\theta)\}_\theta \cup \{R_Z(\theta)X\}_\theta \cong \mathrm{O}(2).
  \label{eq:delocal-group}
\end{equation}
This family is non-universal, so a general single-qubit program eventually needs a gate that couples the rails.

For a single dual-rail qubit, the loss model excludes completions that transfer population from the vacuum $|00\rangle$ to $|11\rangle$, because a later photon-loss event can then return the erased state to the code space (seepage)~\cite{wood_leakage_2018}. An off-diagonal $A$ permits this transfer, so an admissible completion must preserve the rail pair's total excitation number, which requires
\begin{equation}
  A=e^{i\alpha}R_z(\beta).
  \label{eq:loss-compatible-complement}
\end{equation}
Varying $\alpha$ moves the completion along the orange line in Fig.~\ref{fig:weyl-sweep}, and varying $\beta$ changes the admissible completion without changing its Weyl class. Because Eq.~\eqref{eq:loss-compatible-complement} makes $A$ diagonal, an admissible completion is a product unitary only when $B$ is diagonal and $\det A=\det B$, so $R_Z(\theta)$ is the only family in Eq.~\eqref{eq:delocal-group} with admissible product completions.

For logical $X$, $\mathcal{P}(X)$ contains every $M(A,X)$, while $\mathcal{A}(X)$ retains the completions whose Weyl classes trace the parametric-SWAP edge from $\mathrm{iSWAP}$ to $\mathrm{SWAP}$~\cite{crooks_gates_2026}. The dual-rail cavity ISA includes the fractional-SWAP family $\mathrm{SWAP}^t$ with $M(I,X^t)=\mathrm{SWAP}^t$, so $\mathcal{R}(X)$ contains the native circuit for the $\mathrm{SWAP}$ endpoint $M(I,X)$. More generally, the encoding maps logical $R_X(\theta)$ to the physical fractional-SWAP rotation $R_{\mathrm{SWAP}}(\theta)$, which couples the two rails, and $\mathcal{R}(R_Z(\theta))$ contains one single-rail rotation.

\subsection{Dual-Rail Two-Qubit Completions}\label{sec:two-qubit-completions}

Let $V_{AB}=V_A\otimes V_B$ encode two logical qubits in four rails. A four-rail unitary $U$ completes logical $\mathrm{CR}_z(\theta)$ when
\begin{equation}
  U V_{AB}=V_{AB}\mathrm{CR}_z(\theta),
  \label{eq:two-qubit-completion}
\end{equation}
where we use the convention
\begin{equation}
  \mathrm{CR}_z(\theta)
  =
  |0\rangle\!\langle0|\otimes I
  +
  |1\rangle\!\langle1|\otimes R_z(\theta).
  \label{eq:crz-convention}
\end{equation}
For the loss model considered here, diagonal completions form an admissible family because they preserve the occupation of every rail and cannot return an erased state to the code space. We construct realizations only from this family without characterizing the full admissible set. Let logical qubit $A$ occupy rails $(a_0,a_1)$ and logical qubit $B$ occupy rails $(b_0,b_1)$. If a diagonal completion uses one interaction between selected rails $a_i$ and $b_j$, its nonlocal action is a controlled phase up to local $R_z$ rotations. The realization set holds four such circuits, one per rail pairing $i,j\in\{0,1\}$. Each uses one physical $\mathrm{CR}_z$ between $a_i$ and $b_j$ and, when $i=1$, one single-rail $R_z$. With $s_j=(-1)^j$, the circuits are
\begin{equation}
  C_{ij}(\theta)
  =
  R_{z,b_j}\!\left(\mathbf{1}_{i=1}s_j\theta\right)
  \mathrm{CR}_{z,a_i b_j}\!\left((-1)^{i+j}\theta\right).
  \label{eq:four-crz-completions}
\end{equation}
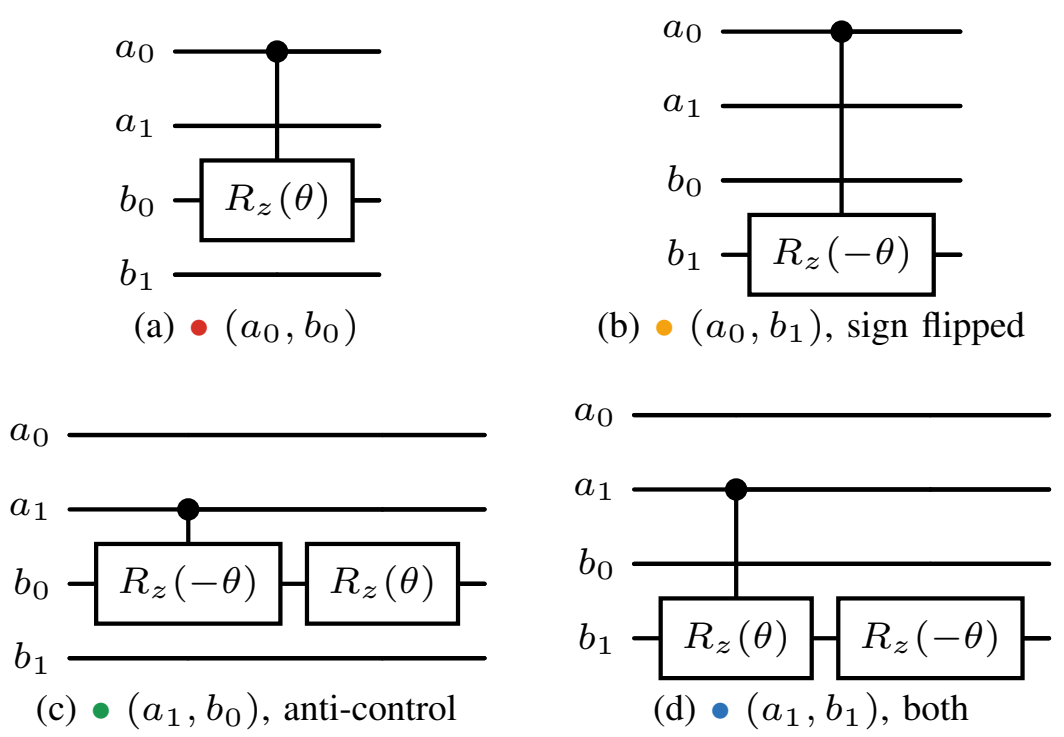
\begin{figure}[t]
  \centering
  \definecolor{beamone}{RGB}{215,48,39}
\definecolor{beamtwo}{RGB}{244,162,15}
\definecolor{beamthree}{RGB}{26,152,80}
\definecolor{beamfour}{RGB}{44,114,184}
\scriptsize
\setlength{\tabcolsep}{6pt}
\begin{tabular}{cc}
\begin{quantikz}[row sep={14pt,between origins}, column sep=5pt]
\lstick{$a_0$} & \ctrl{2}           & \qw \\
\lstick{$a_1$} & \qw                & \qw \\
\lstick{$b_0$} & \gate{R_z(\theta)} & \qw \\
\lstick{$b_1$} & \qw                & \qw
\end{quantikz}
&
\begin{quantikz}[row sep={14pt,between origins}, column sep=5pt]
\lstick{$a_0$} & \ctrl{3}            & \qw \\
\lstick{$a_1$} & \qw                 & \qw \\
\lstick{$b_0$} & \qw                 & \qw \\
\lstick{$b_1$} & \gate{R_z(-\theta)} & \qw
\end{quantikz}
\\[2pt]
(a) \textcolor{beamone}{$\bullet$} $(a_0, b_0)$ & (b) \textcolor{beamtwo}{$\bullet$} $(a_0, b_1)$, sign flipped \\[8pt]
\begin{quantikz}[row sep={14pt,between origins}, column sep=5pt]
\lstick{$a_0$} & \qw                 & \qw                & \qw \\
\lstick{$a_1$} & \ctrl{1}            & \qw                & \qw \\
\lstick{$b_0$} & \gate{R_z(-\theta)} & \gate{R_z(\theta)} & \qw \\
\lstick{$b_1$} & \qw                 & \qw                & \qw
\end{quantikz}
&
\begin{quantikz}[row sep={14pt,between origins}, column sep=5pt]
\lstick{$a_0$} & \qw                & \qw                 & \qw \\
\lstick{$a_1$} & \ctrl{2}           & \qw                 & \qw \\
\lstick{$b_0$} & \qw                & \qw                 & \qw \\
\lstick{$b_1$} & \gate{R_z(\theta)} & \gate{R_z(-\theta)} & \qw
\end{quantikz}
\\[2pt]
(c) \textcolor{beamthree}{$\bullet$} $(a_1, b_0)$, anti-control & (d) \textcolor{beamfour}{$\bullet$} $(a_1, b_1)$, both \\
\end{tabular}

  \caption{Each of the four realizations of logical dual-rail
    $\mathrm{CR}_z(\theta)$ uses one rail pairing, which fixes the rotation
    sign and the single-rail $R_z$ correction.}
  \label{fig:crz-candidates}
\end{figure}

Choosing the other control rail inverts the control condition, while choosing the other target rail reverses the controlled-rotation angle in Eq.~\eqref{eq:four-crz-completions}. An inverted control has the equivalent forms
\begin{equation}
  \scalebox{0.84}{$\vcenter{\hbox{%
          \begin{quantikz}[row sep={15pt,between origins}, column sep=4pt]
            \lstick{$c$} & \octrl{1}          & \\
            \lstick{$t$} & \gate{R_z(\theta)} &
          \end{quantikz}}}$}
  ={}
  \scalebox{0.84}{$\vcenter{\hbox{%
          \begin{quantikz}[row sep={15pt,between origins}, column sep=4pt]
            \lstick{$c$} & \gate{X} & \ctrl{1}           & \gate{X} & \\
            \lstick{$t$} &          & \gate{R_z(\theta)} &          &
          \end{quantikz}}}$}
  ={}
  \scalebox{0.84}{$\vcenter{\hbox{%
          \begin{quantikz}[row sep={15pt,between origins}, column sep=4pt]
            \lstick{$c$} & \ctrl{1}                   & & \\
            \lstick{$t$} & \gate{R_z(\text{-}\theta)} & \gate{R_z(\theta)} &
          \end{quantikz}}}$}
  \label{eq:open-control}
\end{equation}
The four circuits use the diagonal form on the right, because the two single-rail $X$ gates of the middle form leave the dual-rail code space.

We implement the physical $\mathrm{CR}_z$ in Eq.~\eqref{eq:four-crz-completions} using the SWS sequence, which produces a controlled phase between two selected cavity modes. Its wait time sets the phase, and calibrated local phase corrections produce CZ~\cite{mehta_bias-preserving_2025}. The controlled phase satisfies
\begin{equation}
  \operatorname{CP}(\theta)
  = \operatorname{diag}(1,1,1,e^{i\theta})
  =
  e^{i\theta/4}
  \bigl(R_z(\theta/2)\otimes I\bigr)
  \mathrm{CR}_z(\theta),
  \label{eq:cp-crz-convention}
\end{equation}
where the $R_z(\theta/2)$ rotation acts on the control. We absorb this local phase correction and the calibrated SWS phase corrections into the physical $\mathrm{CR}_z$ instruction.

A direct calculation on the four logical basis states gives $C_{ij}(\theta)V_{AB}=V_{AB}\mathrm{CR}_z(\theta)$. The normalized physical $\mathrm{CR}_z$ and every explicit gate in $C_{ij}$ are diagonal at the target level, so each circuit preserves both pairs' occupation and is admissible.

The compiler therefore exposes the logical ISA $\{R_X,R_Z,\mathrm{CR}_z\}$, with $R_X(\theta)$ mapped to $R_{\mathrm{SWAP}}(\theta)$, $R_Z(\theta)$ to a single-rail rotation, and $\mathrm{CR}_z(\theta)$ to the four realizations of Eq.~\eqref{eq:four-crz-completions}.

\subsection{Dual-Rail Logical Availability as a Quotient}\label{sec:quotient}

With the four realizations of Eq.~\eqref{eq:four-crz-completions}, logical $\mathrm{CR}_z$ is available on two logical qubits exactly when a device edge connects a rail of one to a rail of the other. The logical availability graph is therefore the simple quotient of the occupied device graph by the rail pairs. Let $S_q(\lambda)$ contain the two vertices occupied by logical qubit $q$, and let $G_\lambda=G[\bigcup_q S_q(\lambda)]$ be the subgraph induced by the occupied vertices. The sets $S_q(\lambda)$ form the partition $\Pi_\lambda$, and the quotient multigraph $Q(G_\lambda,\Pi_\lambda)$ has one vertex for each logical qubit, an edge for each device edge between two rail pairs, and a self-loop for each device edge within one rail pair. Removing self-loops and merging parallel edges gives $H_{\mathrm{CR}_z}(G,\lambda)$. This construction holds only for realizations with one interaction. For richer realization sets, such as the aligned logical SWAP of Section~\ref{sec:baselines} (Fig.~\ref{fig:logical-swap-candidates}a), which needs two disjoint edges, Eq.~\eqref{eq:availability-relation} still defines availability but the quotient does not compute it.

Representing the two rails as one logical vertex combines their physical neighborhoods, so on a device of maximum degree $\Delta$ the logical vertex has at most $2\Delta$ distinct neighbors. An edge between the rails becomes a discarded self-loop, so separating the rails can expose more external neighbors. In Fig.~\ref{fig:availability-graph}, the paired layout $\lambda_0$ gives the path $P_6$ in panels (a) and (b). The SWAPs shown produce the logical availability graphs $K_{3,3}$ in panels (c) and (d) and then $K_6$ in panels (e) and (f).

\section{Compiler Design}\label{sec:compiler-design}
\begin{figure}[t]
  \centering
    \begin{tikzpicture}[
    x=1cm,
    y=1cm,
    font=\footnotesize,
    flow/.style={-{Stealth[length=3pt]}, line width=0.65pt},
    dep/.style={-{Stealth[length=2.6pt]}, line width=0.55pt},
    route/.style={
        draw=black!55,
        fill=black!2,
        rounded corners=7pt,
        line width=0.65pt,
        minimum height=7.2mm,
        align=center,
        inner xsep=4pt,
        inner ysep=1.5pt
      },
    setnode/.style={
        draw=prismaccent,
        fill=prismaccent!6,
        rounded corners=2pt,
        line width=0.8pt,
        minimum width=1.05cm,
        minimum height=6.6mm,
        align=center,
        inner sep=1pt
      },
    fixed/.style={
        circle,
        draw=black!60,
        fill=white,
        line width=0.6pt,
        minimum size=5.4mm,
        inner sep=0.5pt
      },
    stage/.style={
        draw=black!55,
        fill=black!2,
        rounded corners=2pt,
        line width=0.6pt,
        minimum height=6.5mm,
        align=center,
        inner xsep=3pt,
        inner ysep=1.5pt
      },
    decision/.style={
        diamond,
        draw=black!60,
        fill=white,
        line width=0.6pt,
        aspect=2.0,
        align=center,
        inner xsep=1pt,
        inner ysep=1pt
      }
    ]
    \definecolor{prismaccent}{RGB}{106,61,154}
    \definecolor{beamone}{RGB}{215,48,39}
    \definecolor{beamtwo}{RGB}{244,162,15}
    \definecolor{beamthree}{RGB}{26,152,80}
    \definecolor{beamfour}{RGB}{44,114,184}

    \node[font=\footnotesize\bfseries, anchor=west] at (0,1.72)
    {(a) Compilers};
    \begin{scope}[xshift=-0.16cm]
      \path[draw=prismaccent, fill=prismaccent!10, line width=0.8pt]
      (3.30,1.30) -- (3.46,1.38) -- (3.46,1.22) -- cycle;
      \node[font=\scriptsize, text=prismaccent, anchor=west]
      at (3.52,1.30) {lowering};
      \path[
        draw=beamthree!85!black,
        fill=beamthree!80,
        line width=0.7pt
      ] (6.30,1.22) -- (6.30,1.38) -- (6.46,1.30) -- cycle;
      \node[font=\scriptsize, text=beamthree!75!black, anchor=west]
      at (6.52,1.30) {commitment};

      \node[anchor=east, align=right] at (1.42,0.55) {Logical-\\SABRE};
      \node[route, minimum width=1.45cm] (lr) at (2.24,0.55) {
        $\circlearrowleft$ SABRE\\[-1pt]
        \scriptsize logical gates
      };
      \path[draw=prismaccent, fill=prismaccent!10, line width=0.8pt]
      (3.12,0.55) -- (3.64,0.81) -- (3.64,0.29) -- cycle;
      \node[setnode] (ls) at (4.32,0.55) {$\mathcal R(g)$};
      \path[
        draw=beamthree!85!black,
        fill=beamthree!80,
        line width=0.7pt
      ] (4.96,0.29) -- (4.96,0.81) -- (5.42,0.55) -- cycle;
      \node[route, minimum width=2.12cm] (lp) at (6.78,0.55) {
        $\circlearrowleft$ SABRE search\\[-1pt]
        \scriptsize physical gates
      };
      \draw[flow] (lr) -- (3.12,0.55);
      \draw[flow] (3.64,0.55) -- (ls);
      \draw[line width=0.65pt] (ls.east) -- (4.96,0.55);
      \draw[flow] (5.42,0.55) -- (lp);

      \node[anchor=east] at (1.42,-0.60) {Greedy};
      \draw[flow] (1.62,-0.60) -- (3.12,-0.60);
      \path[draw=prismaccent, fill=prismaccent!10, line width=0.8pt]
      (3.12,-0.60) -- (3.64,-0.34) -- (3.64,-0.86) -- cycle;
      \node[setnode] (gs) at (4.32,-0.60) {$\mathcal R(g)$};
      \node[route, minimum width=3.10cm] (gr) at (6.72,-0.60) {};
      \node[align=center] at ([xshift=0.29cm]gr.center) {
        $\circlearrowleft$ SABRE search\\[-1pt]
        \scriptsize committed realizations
      };
      \draw[flow] (3.64,-0.60) -- (gs);
      \draw[line width=0.65pt] (gs.east) -- (5.32,-0.60);
      \path[
        draw=beamthree!85!black,
        fill=beamthree!80,
        line width=0.7pt
      ] (5.32,-0.86) -- (5.32,-0.34) -- (5.78,-0.60) -- cycle;

      \node[anchor=east] at (1.42,-1.75) {\sysname{}};
      \draw[flow] (1.62,-1.75) -- (3.12,-1.75);
      \path[draw=prismaccent, fill=prismaccent!10, line width=0.8pt]
      (3.12,-1.75) -- (3.64,-1.49) -- (3.64,-2.01) -- cycle;
      \node[setnode] (ps) at (4.32,-1.75) {$\mathcal R(g)$};
      \node[
        route,
        draw=prismaccent,
        fill=prismaccent!4,
        minimum width=2.50cm
      ] (pr) at (6.42,-1.75) {
        $\circlearrowleft$ SABRE search\\[-1pt]
        \scriptsize realization sets
      };
      \draw[flow] (3.64,-1.75) -- (ps);
      \draw[flow] (ps) -- (pr);
      \draw[line width=0.65pt] (pr.east) -- (7.78,-1.75);
      \path[
        draw=beamthree!85!black,
        fill=beamthree!80,
        line width=0.7pt
      ] (7.78,-2.01) -- (7.78,-1.49) -- (8.24,-1.75) -- cycle;

      \foreach \target in {ls,gs,ps} {
          \foreach \cc/\xx in
            {beamone/-0.18,beamtwo/-0.06,beamthree/0.06,beamfour/0.18} {
              \fill[\cc] ([xshift=\xx cm,yshift=2.2pt]\target.south)
              circle (1.0pt);
            }
        }
    \end{scope}

    \draw[black!20, line width=0.5pt] (0,-2.34) -- (8.30,-2.34);
    \node[font=\footnotesize\bfseries, anchor=west] at (0,-2.72)
    {(b) SABRE loop over the dependency graph};

    \node[font=\scriptsize\bfseries, text=prismaccent] at (0.42,-3.28)
    {front $F$};
    \draw[prismaccent, line width=0.8pt] (0.06,-3.42) -- (0.78,-3.42);
    \node[font=\scriptsize\bfseries, text=black!55] at (1.78,-3.28)
    {lookahead $F'$};
    \draw[black!35, line width=0.8pt] (1.20,-3.42) -- (2.36,-3.42);

    \node[fixed] (f1) at (0.62,-4.05) {$f$};
    \node[setnode] (fn) at (0.62,-5.05) {$g$};
    \foreach \cc/\xx in
      {beamone/-0.18,beamtwo/-0.06,beamthree/0.06,beamfour/0.18} {
        \fill[\cc] ([xshift=\xx cm,yshift=2.5pt]fn.south) circle (1.0pt);
      }
    \node[fixed] (x1) at (1.72,-4.05) {$g'$};
    \node[fixed] (x2) at (1.72,-5.05) {$g''$};
    \draw[dep] (f1) -- (x1);
    \draw[dep] (fn) -- (x1);
    \draw[dep] (fn) -- (x2);

    \draw[
      black!35,
      dashed,
      rounded corners=7pt,
      line width=0.6pt
    ] (2.74,-5.76) rectangle (6.80,-3.45);
    \node[
      font=\scriptsize\bfseries,
      text=black!60,
      anchor=west,
      fill=white,
      inner xsep=2pt
    ] at (2.98,-3.45) {SABRE};

    \node[
      stage,
      font=\scriptsize,
      minimum width=1.22cm
    ] (score) at (3.75,-4.53)
    {score $F,F'$};
    \foreach \cc/\xx in
      {beamone/-0.18,beamtwo/-0.06,beamthree/0.06,beamfour/0.18} {
        \fill[\cc] ([xshift=\xx cm,yshift=2.0pt]score.south) circle (0.9pt);
      }
    \node[
      decision,
      font=\scriptsize,
      minimum width=1.24cm,
      minimum height=6.8mm
    ] (ready) at (5.43,-4.53)
    {executable?};
    \node[
      stage,
      font=\scriptsize,
      minimum width=1.00cm,
      minimum height=5.4mm,
      inner ysep=0.6pt
    ] (swap) at (5.43,-5.42)
    {SWAP};
    \node[
      stage,
      draw=beamthree!85!black,
      fill=beamthree!7,
      font=\scriptsize,
      minimum width=7.0mm
    ] (emit) at (7.93,-4.53)
    {emit};

    \draw[flow] (2.12,-4.53) -- (score.west);
    \draw[flow] (score) -- (ready);
    \draw[line width=0.65pt] (ready.east) --
    node[above, font=\scriptsize] {yes}
    (6.94,-4.53);
    \path[
      draw=beamthree!85!black,
      fill=beamthree!80,
      line width=0.7pt
    ] (6.94,-4.78) -- (6.94,-4.28) -- (7.32,-4.53) -- cycle;
    \draw[flow] (7.32,-4.53) -- (emit);
    \draw[flow] (ready.south) -- (swap.north);
    \node[font=\scriptsize, anchor=north west, inner ysep=1pt] at ([xshift=1pt]ready.south) {no};
    \draw[flow] (swap.west) -| ([xshift=-0.42cm]score.south);
    \node[font=\scriptsize, text=black!55, anchor=south] at (4.14,-5.42)
    {new $\lambda$};
  \end{tikzpicture}
  \caption{(a) Each purple prism is the logical--physical boundary, where \textsc{LowerExpansion} expands a logical gate into its realization set $\mathcal{R}(g)$ (colored dots), and each green arrowhead marks commitment to one realization. Logical-SABRE commits at lowering, Greedy before each router call, and \sysname{} when each gate executes. (b) The router scores each unresolved gate by its nearest realization, commits executable gates in the front layer, and otherwise inserts the best-scoring SWAP.}
  \label{fig:compiler-strategies}
\end{figure}
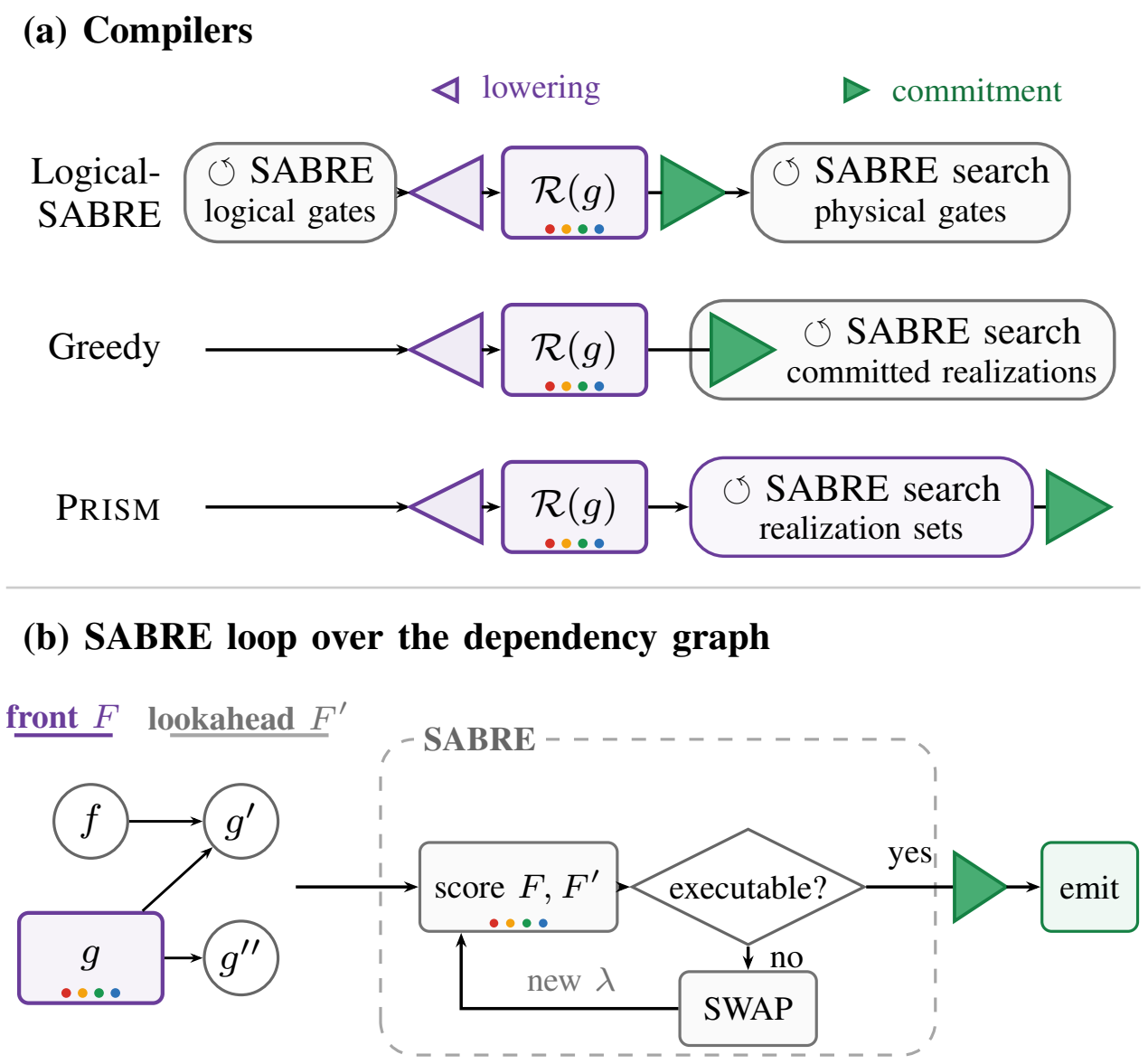

\sysname{} carries each logical gate across the logical--physical boundary as a set of realizations and commits to one when routing reaches a layout in which that realization can execute. Greedy commits before each router call, and Logical-SABRE after routing on a fixed logical coupling graph (Fig.~\ref{fig:compiler-strategies}a).

\subsection{Boundary Intermediate Representation}\label{sec:ir}

\textsc{LowerExpansion} produces the boundary IR that \sysname{} and Greedy share. It replaces each logical qubit with the wires that encode it and each logical gate with a \textsc{LoweredOp} over those wires. A \textsc{LoweredOp} stores the realization set and depends on every wire any realization uses until commitment.

A logical $\mathrm{CR}_z$ on $(a_0,a_1)$ and $(b_0,b_1)$ becomes one \textsc{LoweredOp} containing the four circuits from Eq.~\eqref{eq:four-crz-completions}, with interaction patterns
\begin{equation}
  \{(a_i,b_j):i,j\in\{0,1\}\}.
  \label{eq:realization-2q}
\end{equation}
At the circuit's start and at each logical reset, \textsc{LowerExpansion} emits a two-rail preparation of $|0_L\rangle$ per logical qubit, which the router must execute on adjacent sites. Measurement reads each rail separately and therefore needs no adjacency.

\subsection{Commitment at Execution}\label{sec:dr-sabre}

\begin{algorithm}[t]
  \caption{\sysname{}'s inner SABRE routing loop.}
  \label{alg:dr-sabre}
  	  \begin{algorithmic}[1]
	    \Function{Route}{$D_{\text{phys}}, G, \lambda_0$}
	    \State $\lambda \gets \lambda_0$;\ \ $\nu \gets 0$ on every site of $G$
	    \While{\textsc{FrontLayer}($D_{\text{phys}}$) is non-empty}
	    \State $F \gets$ \textsc{FrontLayer}($D_{\text{phys}}$);\ \ progress $\gets$ \textsc{false}
	    \For{each $g \in F$}
	    \State $\mathcal{R}_\lambda(g) \gets \{r \in \mathcal{R}(g):\delta_\lambda(r)\le 1\}$
	    \If{$\mathcal{R}_\lambda(g)$ is non-empty}
	    \State $r^* \gets \textsc{LeastUsed}(\mathcal{R}_\lambda(g))$
	    \State \textsc{Emit}$(C_{r^*})$;\ \ \textsc{Remove}$(g,\,D_{\text{phys}})$
	    \State $\nu \gets \nu+1$ at both sites of $r^*$ if $\smash{|\mathcal{R}(g)|}>1$
	    \State progress $\gets$ \textsc{true}
	    \EndIf
	    \EndFor
	    \If{\textbf{not} progress}
	    \State $\nu \gets 0.9\,\nu$ if gates followed the last SWAP
	    \State $F' \gets \textsc{Lookahead}(D_{\text{phys}},F)$
	    \State $s^* \gets \textsc{SabreSwap}(F,F',G,\lambda,h_\lambda)$
	    \State \textsc{Emit}$(s^*)$;\quad $\lambda\gets\operatorname{swap}(\lambda,s^*)$
	    \EndIf
	    \EndWhile
	    \State \Return the emitted operations and $\lambda$
	    \EndFunction
  \end{algorithmic}

\end{algorithm}

Algorithm~\ref{alg:dr-sabre} routes the directed acyclic graph (DAG) $D_{\text{phys}}$ of lowered operations. Each \textsc{LoweredOp} remains unresolved until the current layout $\lambda$ supports a realization. For a one-interaction realization $r$ on wires $(u_r,v_r)$, let
\begin{equation}
  \delta_\lambda(r)=d_G(\lambda(u_r),\lambda(v_r)),
  \label{eq:realization-distance}
\end{equation}
where $d_G$ is the shortest-path distance on the device graph $G$. A realization with no interaction, such as a single-rail rotation, has $\delta_\lambda(r)=0$. The routing cost of an unresolved gate $g$ is
\begin{equation}
  h_\lambda(g)=\min_{r\in\mathcal{R}(g)}\delta_\lambda(r).
  \label{eq:prism-routing-cost}
\end{equation}
SABRE uses this cost in the front layer and lookahead set (line 17). The favored realization can therefore change with the layout while the rest of SABRE's candidate-SWAP neighborhood, weighting, and decay rules remain unchanged~\cite{li_tackling_2019}.

An unresolved gate is executable when at least one realization has $\delta_\lambda(r)\le 1$ (line 6), so a realization with no interaction, which has $\delta_\lambda(r)=0$, is always executable. To shorten the schedule, \sysname{} keeps a usage count $\nu$ at each device site and favors less-used sites when choosing among executable realizations. If several realizations are executable, \textsc{LeastUsed} (line 8) commits the one whose two sites have the lowest total count and breaks remaining ties in a fixed order. The router inserts SWAPs until a front-layer gate becomes executable, then executes every gate that becomes executable. If gates have executed since the last SWAP, it multiplies all counts by 0.9 before choosing the next SWAP (line 15). Every forward pass, backward pass, and final trial restores each \textsc{LoweredOp}'s full realization set and resets all counts to zero, so only the layout carries over.

An operation with one realization, such as $R_{\mathrm{SWAP}}$ or preparation, enters the loop as a one-element set without changing usage counts. A backward pass routes the preparations last. If a refined layout separates a rail pair, the next forward or final pass must reunite it before preparing it, and those SWAPs are counted. The lowering pass replaces each committed \textsc{LoweredOp} with its selected circuit.

\subsection{Baseline Compilers}\label{sec:baselines}

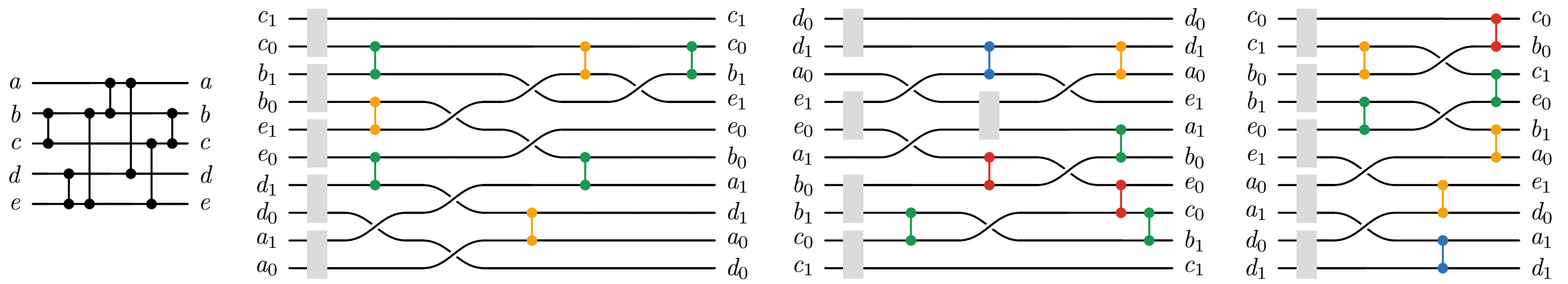
\begin{figure*}[t]
  \centering
  \definecolor{beamone}{RGB}{215,48,39}
\definecolor{beamtwo}{RGB}{244,162,15}
\definecolor{beamthree}{RGB}{26,152,80}
\definecolor{beamfour}{RGB}{44,114,184}
\resizebox{\textwidth}{!}{%
\begin{tabular}{c@{\hskip 0.8em}c@{\hskip 0.8em}c@{\hskip 0.8em}c}
\begin{quantikz}[row sep={12pt,between origins}, column sep=4pt]
  \lstick{$\mathit{a}$} & \qw & \qw & \qw & \ctrl{1} & \ctrl{3} & \qw & \qw & \rstick{$\mathit{a}$} \qw \\
  \lstick{$\mathit{b}$} & \ctrl{1} & \qw & \ctrl{3} & \control{} & \qw & \qw & \ctrl{1} & \rstick{$\mathit{b}$} \qw \\
  \lstick{$\mathit{c}$} & \control{} & \qw & \qw & \qw & \qw & \ctrl{2} & \control{} & \rstick{$\mathit{c}$} \qw \\
  \lstick{$\mathit{d}$} & \qw & \ctrl{1} & \qw & \qw & \control{} & \qw & \qw & \rstick{$\mathit{d}$} \qw \\
  \lstick{$\mathit{e}$} & \qw & \control{} & \control{} & \qw & \qw & \control{} & \qw & \rstick{$\mathit{e}$} \qw
\end{quantikz}
&
\begin{quantikz}[row sep={11pt,between origins}, column sep=7pt]
  \lstick{$\mathit{c}_1$} & \gate[wires=2,style={fill=black!14,draw=none}]{} & \qw & \qw & \qw & \qw & \qw & \qw & \rstick{$\mathit{c}_1$} \qw \\
  \lstick{$\mathit{c}_0$} &  & \control[style={draw=beamthree,fill=beamthree}]{} & \qw & \qw & \ctrl[style={draw=beamtwo,fill=beamtwo},wire style={draw=beamtwo}]{1} & \qw & \control[style={draw=beamthree,fill=beamthree}]{} & \rstick{$\mathit{c}_0$} \qw \\
  \lstick{$\mathit{b}_1$} & \gate[wires=2,style={fill=black!14,draw=none}]{} & \ctrl[style={draw=beamthree,fill=beamthree},wire style={draw=beamthree}]{-1} & \qw & \permute{2,1} & \control[style={draw=beamtwo,fill=beamtwo}]{} & \permute{2,1} & \ctrl[style={draw=beamthree,fill=beamthree},wire style={draw=beamthree}]{-1} & \rstick{$\mathit{b}_1$} \qw \\
  \lstick{$\mathit{b}_0$} &  & \ctrl[style={draw=beamtwo,fill=beamtwo},wire style={draw=beamtwo}]{1} & \permute{2,1} &  & \qw &  & \qw & \rstick{$\mathit{e}_1$} \qw \\
  \lstick{$\mathit{e}_1$} & \gate[wires=2,style={fill=black!14,draw=none}]{} & \control[style={draw=beamtwo,fill=beamtwo}]{} &  & \permute{2,1} & \qw & \qw & \qw & \rstick{$\mathit{e}_0$} \qw \\
  \lstick{$\mathit{e}_0$} &  & \control[style={draw=beamthree,fill=beamthree}]{} & \qw &  & \control[style={draw=beamthree,fill=beamthree}]{} & \qw & \qw & \rstick{$\mathit{b}_0$} \qw \\
  \lstick{$\mathit{d}_1$} & \gate[wires=2,style={fill=black!14,draw=none}]{} & \ctrl[style={draw=beamthree,fill=beamthree},wire style={draw=beamthree}]{-1} & \permute{2,1} & \qw & \ctrl[style={draw=beamthree,fill=beamthree},wire style={draw=beamthree}]{-1} & \qw & \qw & \rstick{$\mathit{a}_1$} \qw \\
  \lstick{$\mathit{d}_0$} &  & \permute{2,1} &  & \control[style={draw=beamtwo,fill=beamtwo}]{} & \qw & \qw & \qw & \rstick{$\mathit{d}_1$} \qw \\
  \lstick{$\mathit{a}_1$} & \gate[wires=2,style={fill=black!14,draw=none}]{} &  & \permute{2,1} & \ctrl[style={draw=beamtwo,fill=beamtwo},wire style={draw=beamtwo}]{-1} & \qw & \qw & \qw & \rstick{$\mathit{a}_0$} \qw \\
  \lstick{$\mathit{a}_0$} &  & \qw &  & \qw & \qw & \qw & \qw & \rstick{$\mathit{d}_0$} \qw
\end{quantikz}
&
\begin{quantikz}[row sep={11pt,between origins}, column sep=7pt]
  \lstick{$\mathit{d}_0$} & \gate[wires=2,style={fill=black!14,draw=none}]{} & \qw & \qw & \qw & \qw & \qw & \rstick{$\mathit{d}_0$} \qw \\
  \lstick{$\mathit{d}_1$} &  & \qw & \ctrl[style={draw=beamfour,fill=beamfour},wire style={draw=beamfour}]{1} & \qw & \control[style={draw=beamtwo,fill=beamtwo}]{} & \qw & \rstick{$\mathit{d}_1$} \qw \\
  \lstick{$\mathit{a}_0$} & \qw & \permute{2,1} & \control[style={draw=beamfour,fill=beamfour}]{} & \permute{2,1} & \ctrl[style={draw=beamtwo,fill=beamtwo},wire style={draw=beamtwo}]{-1} & \qw & \rstick{$\mathit{a}_0$} \qw \\
  \lstick{$\mathit{e}_1$} & \gate[wires=2,style={fill=black!14,draw=none}]{} &  & \gate[wires=2,style={fill=black!14,draw=none}]{} &  & \qw & \qw & \rstick{$\mathit{e}_1$} \qw \\
  \lstick{$\mathit{e}_0$} &  & \permute{2,1} &  & \qw & \ctrl[style={draw=beamthree,fill=beamthree},wire style={draw=beamthree}]{1} & \qw & \rstick{$\mathit{a}_1$} \qw \\
  \lstick{$\mathit{a}_1$} & \qw &  & \control[style={draw=beamone,fill=beamone}]{} & \permute{2,1} & \control[style={draw=beamthree,fill=beamthree}]{} & \qw & \rstick{$\mathit{b}_0$} \qw \\
  \lstick{$\mathit{b}_0$} & \gate[wires=2,style={fill=black!14,draw=none}]{} & \qw & \ctrl[style={draw=beamone,fill=beamone},wire style={draw=beamone}]{-1} &  & \control[style={draw=beamone,fill=beamone}]{} & \qw & \rstick{$\mathit{e}_0$} \qw \\
  \lstick{$\mathit{b}_1$} &  & \ctrl[style={draw=beamthree,fill=beamthree},wire style={draw=beamthree}]{1} & \permute{2,1} & \qw & \ctrl[style={draw=beamone,fill=beamone},wire style={draw=beamone}]{-1} & \control[style={draw=beamthree,fill=beamthree}]{} & \rstick{$\mathit{c}_0$} \qw \\
  \lstick{$\mathit{c}_0$} & \gate[wires=2,style={fill=black!14,draw=none}]{} & \control[style={draw=beamthree,fill=beamthree}]{} &  & \qw & \qw & \ctrl[style={draw=beamthree,fill=beamthree},wire style={draw=beamthree}]{-1} & \rstick{$\mathit{b}_1$} \qw \\
  \lstick{$\mathit{c}_1$} &  & \qw & \qw & \qw & \qw & \qw & \rstick{$\mathit{c}_1$} \qw
\end{quantikz}
&
\begin{quantikz}[row sep={11pt,between origins}, column sep=7pt]
  \lstick{$\mathit{c}_0$} & \gate[wires=2,style={fill=black!14,draw=none}]{} & \qw & \qw & \control[style={draw=beamone,fill=beamone}]{} & \rstick{$\mathit{c}_0$} \qw \\
  \lstick{$\mathit{c}_1$} &  & \control[style={draw=beamtwo,fill=beamtwo}]{} & \permute{2,1} & \ctrl[style={draw=beamone,fill=beamone},wire style={draw=beamone}]{-1} & \rstick{$\mathit{b}_0$} \qw \\
  \lstick{$\mathit{b}_0$} & \gate[wires=2,style={fill=black!14,draw=none}]{} & \ctrl[style={draw=beamtwo,fill=beamtwo},wire style={draw=beamtwo}]{-1} &  & \ctrl[style={draw=beamthree,fill=beamthree},wire style={draw=beamthree}]{1} & \rstick{$\mathit{c}_1$} \qw \\
  \lstick{$\mathit{b}_1$} &  & \ctrl[style={draw=beamthree,fill=beamthree},wire style={draw=beamthree}]{1} & \permute{2,1} & \control[style={draw=beamthree,fill=beamthree}]{} & \rstick{$\mathit{e}_0$} \qw \\
  \lstick{$\mathit{e}_0$} & \gate[wires=2,style={fill=black!14,draw=none}]{} & \control[style={draw=beamthree,fill=beamthree}]{} &  & \control[style={draw=beamtwo,fill=beamtwo}]{} & \rstick{$\mathit{b}_1$} \qw \\
  \lstick{$\mathit{e}_1$} &  & \permute{2,1} & \qw & \ctrl[style={draw=beamtwo,fill=beamtwo},wire style={draw=beamtwo}]{-1} & \rstick{$\mathit{a}_0$} \qw \\
  \lstick{$\mathit{a}_0$} & \gate[wires=2,style={fill=black!14,draw=none}]{} &  & \control[style={draw=beamtwo,fill=beamtwo}]{} & \qw & \rstick{$\mathit{e}_1$} \qw \\
  \lstick{$\mathit{a}_1$} &  & \permute{2,1} & \ctrl[style={draw=beamtwo,fill=beamtwo},wire style={draw=beamtwo}]{-1} & \qw & \rstick{$\mathit{d}_0$} \qw \\
  \lstick{$\mathit{d}_0$} & \gate[wires=2,style={fill=black!14,draw=none}]{} &  & \ctrl[style={draw=beamfour,fill=beamfour},wire style={draw=beamfour}]{1} & \qw & \rstick{$\mathit{a}_1$} \qw \\
  \lstick{$\mathit{d}_1$} &  & \qw & \control[style={draw=beamfour,fill=beamfour}]{} & \qw & \rstick{$\mathit{d}_1$} \qw
\end{quantikz}
\\[4pt]
(a) logical input & (b) Logical-SABRE (8 SWAPs, 7 drawn, depth 6) & (c) Greedy (5 SWAPs, depth 5) & (d) \sysname{} (4 SWAPs, depth 3) \\
\end{tabular}}

  \caption{Delaying commitment reduces routing SWAPs and depth on a constructed five-qubit example on a ten-site line. Panel (a) is the logical input, and (b)--(d) are the routes of the three compilers. Wires cross at each SWAP and keep their rail labels, and gray bars mark preparation. Colors identify the (control rail, target rail) pairing used for each logical gate in panel (a), as in Fig.~\ref{fig:crz-candidates}. Each column is one layer, with gates placed as early as their wires allow. Depth omits preparation and single-rail phase corrections, and panel (b) shows seven SWAPs because cleanup removes one before measurement (Section~\ref{sec:eval-setup}).}
  \label{fig:strategy-ladder}
\end{figure*}

\paragraph{Greedy}

Before each router call, Greedy selects for every unresolved gate $g$ the nearest realization under its entry layout $\lambda$:
\begin{equation}
  r_\lambda(g)=
  \arg\min_{r\in\mathcal{R}(g)}
  \sum_{(u,v)\in I_r}d_G(\lambda(u),\lambda(v)).
  \label{eq:greedy-policy}
\end{equation}
Greedy breaks ties using the same fixed ordering as \sysname{}. Greedy holds these selections fixed while the router inserts SWAPs, so all final routing trials use the same selections. The entry layout can separate a rail pair that the route has yet to prepare, as it separates the rails of $a$ in Fig.~\ref{fig:strategy-ladder}c.

\paragraph{Logical-SABRE}

Logical-SABRE routes before it lowers. It forms disjoint adjacent rail pairs from a maximum-cardinality matching of the device graph that NetworkX~\cite{hagberg_networkx_2008} returns, which maximizes the number of supported logical qubits and breaks ties among maximum matchings by the order in which the device edges are listed. \textsc{SabreLayout} and \textsc{SabreSwap} then route the logical gates on the logical coupling graph, the simple quotient of the subgraph induced by the matched sites, with one vertex per matched edge and none for an unmatched site. Routing on this graph guarantees at least one available realization for each logical $\mathrm{CR}_z$, and lowering then commits each gate to the first realization whose interactions are all available at the matched sites. Logical-SABRE's search uses the same budget of initial layouts, refinement iterations, and final trials as Greedy and \sysname{} (Section~\ref{sec:eval-setup}), but selects routes by logical SWAP count without accounting for physical SWAPs added later. Its cost depends on its fixed site matching (Section~\ref{sec:eval-ablations}).

\begin{figure}[t]
  \centering
  \scriptsize
\resizebox{\columnwidth}{!}{%
\begin{tabular}{@{}c@{\hskip 14pt}c@{\hskip 14pt}c@{}}
\begin{quantikz}[row sep={13pt,between origins}, column sep=8pt]
\lstick{$a_0$} & \swap{2} & \qw      & \rstick{$b_0$}\qw \\
\lstick{$a_1$} & \qw      & \swap{2} & \rstick{$b_1$}\qw \\
\lstick{$b_0$} & \targX{} & \qw      & \rstick{$a_0$}\qw \\
\lstick{$b_1$} & \qw      & \targX{} & \rstick{$a_1$}\qw
\end{quantikz}
&
\begin{quantikz}[row sep={13pt,between origins}, column sep=8pt]
\lstick{$a_0$} & \swap{3} & \qw      & \swap{1} & \rstick{$b_0$}\qw \\
\lstick{$a_1$} & \qw      & \swap{1} & \targX{} & \rstick{$b_1$}\qw \\
\lstick{$b_0$} & \qw      & \targX{} & \swap{1} & \rstick{$a_0$}\qw \\
\lstick{$b_1$} & \targX{} & \qw      & \targX{} & \rstick{$a_1$}\qw
\end{quantikz}
&
\begin{quantikz}[row sep={13pt,between origins}, column sep=8pt]
\lstick{$a_0$} & \swap{3} & \qw      & \gate[2,style={dashed,dash pattern=on 1.6pt off 1.3pt,fill=none,inner xsep=0.5pt,inner ysep=-2pt,minimum width=0pt}]{X_L} & \rstick{$b_0$}\qw \\
\lstick{$a_1$} & \qw      & \swap{1} &                                        & \rstick{$b_1$}\qw \\
\lstick{$b_0$} & \qw      & \targX{} & \gate[2,style={dashed,dash pattern=on 1.6pt off 1.3pt,fill=none,inner xsep=0.5pt,inner ysep=-2pt,minimum width=0pt}]{X_L} & \rstick{$a_0$}\qw \\
\lstick{$b_1$} & \targX{} & \qw      &                                        & \rstick{$a_1$}\qw
\end{quantikz}
\\[2pt]
(a) aligned &
(b) crossed, corrected &
(c) crossed, frames
\end{tabular}
}

  \caption{Physical realizations of a logical dual-rail SWAP. (a) Two
    aligned SWAPs move each rail to the same-index rail of the other
    pair. (b) Two crossed SWAPs leave each pair role-reversed, and two
    intra-pair SWAPs restore the roles. (c) The dashed $X_L$ boxes are
    frames tracked in software.}
  \label{fig:logical-swap-candidates}
\end{figure}
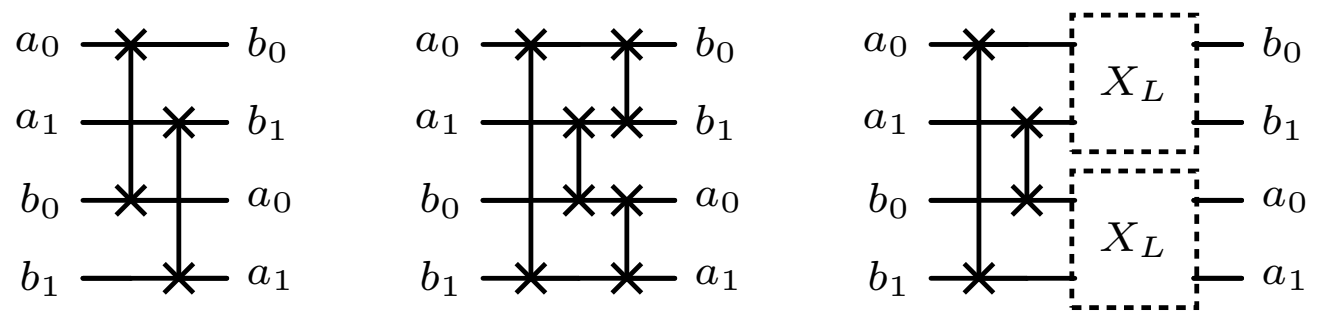

Unlike Greedy and \sysname{}, which insert only physical SWAPs, Logical-SABRE inserts logical SWAPs, which move both rails of each qubit even when moving one would make a gate executable. Each logical SWAP lowers to one of the circuits in Fig.~\ref{fig:logical-swap-candidates}, either two aligned SWAPs (a) or two crossed SWAPs followed by two intra-pair SWAPs (b) that remove the $X_L$ the crossing leaves on each qubit. Circuit (c) would instead track that $X_L$ as a Pauli frame, which Logical-SABRE does not do. Circuits (a) and (b) both need two disjoint device edges between the rail pairs, which with the two matched edges form a four-cycle. Logical-SABRE commits to the first of (a) and (b) whose interactions are available at the matched sites and otherwise commits to (a) and leaves it to Qiskit's stock \textsc{SabreSwap} on the device graph, so on a device graph without four-cycles every logical SWAP takes this fallback. We count two physical SWAPs for circuit (a), four for (b), and any SWAPs added by physical routing. When these SWAPs move rails away from their matched sites, the router routes the previously chosen $\mathrm{CR}_z$ circuits as fixed two-qubit gates.

Figure~\ref{fig:strategy-ladder} compares the three compilers on a constructed example using Qiskit's stock SABRE search budget of 20 random initial layouts, four refinement iterations, and 20 final trials. Logical-SABRE uses eight routing SWAPs at depth six, Greedy five at depth five, and \sysname{} four at depth three.

\section{Evaluation}\label{sec:eval}

\begin{figure*}[t]
  \centering
  \includegraphics[width=\textwidth]{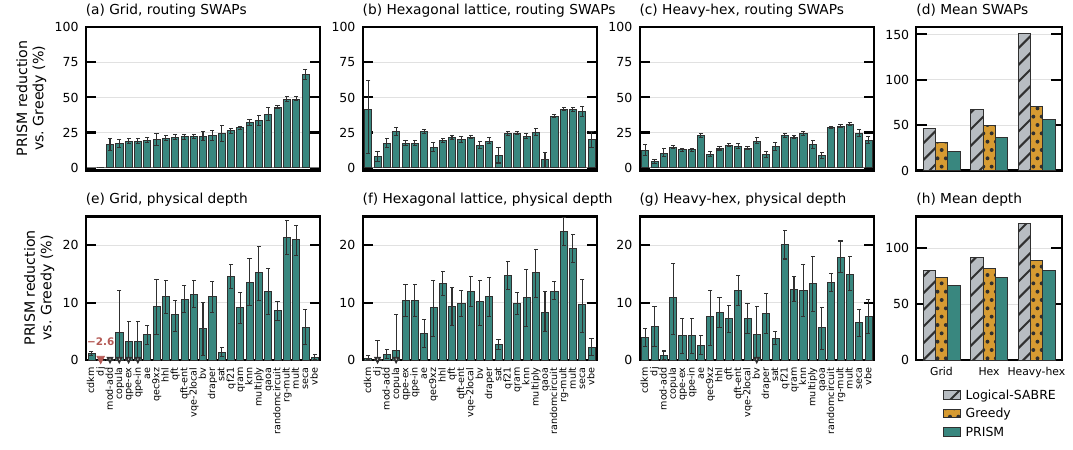}
  \caption{Panels (a) to (c) and (e) to (g) show \sysname{}'s SWAP and depth reductions relative to Greedy on each device graph. Bars compare 64-seed means with 95\% paired seed-bootstrap intervals. Gray triangles mark intervals that extend below the axis, the red triangle gives the one negative mean (dj on the grid, $-2.6\%$ depth), and SWAP bars are omitted where Greedy averages under one routing SWAP (vbe and cdkm on the grid). Panels (d) and (h) show each compiler's mean routing SWAPs and physical depth, over the 24 workloads that Logical-SABRE can place on heavy-hex.}
  \label{fig:eval-core-benchmarks}
\end{figure*}

\subsection{Experimental Setup}\label{sec:eval-setup}

Seventeen workloads come from MQT Bench 2.2.2 at eight logical qubits, or seven for the VBE adder, which has no eight-qubit instance~\cite{quetschlich_mqtbench_2023}. We exclude nine other MQT workloads on which Greedy and \sysname{} insert at most four routing SWAPs in every run on each device graph. The other eight workloads are the largest circuit of each QASMBench medium-scale family that needs routing, at its native width of 11 to 25 logical qubits~\cite{li_qasmbench_2023}. A workload of $n$ logical qubits runs on a grid, a hexagonal lattice, and a heavy-hex graph~\cite{chamberland_topological_2020}, each with at least $2n+2$ sites, which hold its rail pairs and one spare pair (Appendix~\ref{app:device-graphs}). The primary experiment averages results over 64 compiler seeds, and error bars and $\pm$ values are 95\% paired-bootstrap confidence intervals over seeds. The scaling and hardware experiments drop randomcircuit because it is a sampled random circuit and use the other 16 MQT workloads (Table~\ref{tab:workloads}) on the grid and hexagonal lattice, plus heavy-hex for hardware, and average results over five seeds.

We implement \textsc{LowerExpansion} as a pass in Qiskit 2.4.1~\cite{javadiabhari_qiskit_2024}. Before it, a logical stage shared by the three dual-rail compilers decomposes multi-controlled gates with one clean ancilla, applies Toffoli, Pauli-frame, and two Hadamard-reduction rewrites, translates the circuit to $R_X$, $R_Z$, and $\mathrm{CR}_z$, absorbs input SWAPs into the wire permutation, and fixes the operand order of each $\mathrm{CR}_z$ and thereby its SWS control. Our router is a fork of Qiskit 2.3.1's Rust LightSABRE router extended to process \textsc{LoweredOp} interactions. Greedy and \sysname{} use the same modified router, downstream physical compiler, stock SABRE heuristic, set of initial layouts, seed schedule, and search budget.

Following LightSABRE's randomized search procedure~\cite{zou_lightsabre_2024}, each compilation tests 40 random initial layouts and the dense, identity, and reverse layouts of Qiskit's SabreLayout. Each layout receives 16 forward--backward refinement iterations, followed by a 40-trial final route. The compiler selects the completed route with the fewest routing SWAPs, using physical depth and then active gate count to break ties. A critical-path list scheduler then reorders commuting $R_z$ and $\mathrm{CR}_z$ gates without changing the SWAPs and keeps each preparation before every gate on its rails. Under the SWAP objective it reschedules the selected route. Under the depth and duration objectives of Section~\ref{sec:eval-ablations} it reschedules every layout's route before selection.

We evaluate each final circuit by routing SWAP count, physical depth, critical-path duration, and compilation time. All three compilers end with a cleanup pass that removes SWAPs directly preceding a measurement, and routing SWAPs are counted before it. Logical $X$ gates, implemented as intra-pair SWAPs, are not counted as routing SWAPs, and software-tracked $Z$ rotations do not count toward depth. For each additive metric we report the reduction in the sum of workload means and the mean per-workload reduction, which weights workloads equally and omits those where Greedy's value is zero.

\subsection{Delayed Commitment Reduces Routing Cost}
\label{sec:eval-core}

\sysname{} reduces Greedy's total routing SWAP count by 30.8\% [30.4, 31.3] on the grid, 26.0\% [25.6, 26.5] on the hexagonal lattice, and 20.4\% [20.1, 20.8] on heavy-hex, or by 30.0\% [26.9, 31.1], 23.1\% [21.7, 24.3], and 17.2\% [16.9, 17.6] with equal workload weights, where brackets give 95\% paired seed-bootstrap intervals. It reduces physical depth by 8.8\%, 10.0\%, and 10.1\% suite-wide and by 8.2\%, 9.2\%, and 8.6\% with equal weights (Fig.~\ref{fig:eval-core-benchmarks}d, h). \sysname{} uses fewer routing SWAPs in 73 of the 75 workload and device-graph means and ties in two, although 202 of the 4,800 individual runs use more SWAPs than Greedy. Physical depth improves in 74 of the 75 means and rises by 0.2 layers for dj on the grid, where route selection minimizes SWAP count before depth. Relative to Logical-SABRE, Greedy reduces total routing SWAPs by 34.0\% on the grid, 26.8\% on the hexagonal lattice, and 52.9\% on heavy-hex, while \sysname{} reduces them by 54.3\%, 45.8\%, and 62.5\% and physical depth by 16.6\%, 19.7\%, and 34.4\%. On heavy-hex these comparisons cover 24 workloads, because the site matching there has 24 pairs and cannot place knn.

\begin{figure}[t]
  \centering
  \includegraphics[width=\columnwidth]{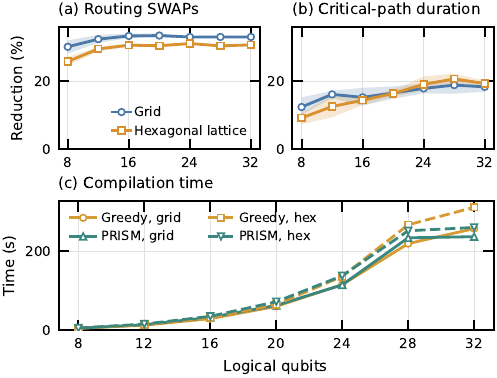}
  \caption{\sysname{} results from 8 to 32 logical qubits. (a), (b) Reductions in routing SWAPs and critical-path duration relative to Greedy. (c) Per-compilation wall time.
    All average results over five seeds for each of the 16 MQT workloads.}
  \label{fig:eval-scaling-impact}
\end{figure}

Across the 16 MQT workloads from eight to 32 logical qubits, \sysname{} uses fewer routing SWAPs than Greedy or ties it for every combination of workload, device graph, and width except two on the hexagonal lattice, one at eight and one at 20 logical qubits. From 12 through 32 logical qubits, it reduces total routing SWAPs by 32.3--33.4\% on the grid and 29.4--31.0\% on the hexagonal lattice (Fig.~\ref{fig:eval-scaling-impact}). At 32 logical qubits the reductions are 33.0\% and 30.6\% (31.8\% and 26.0\% with equal workload weights), and \sysname{} also reduces physical depth by 22.6\% and 23.4\% and critical-path duration by 18.3\% and 19.3\%.

The \sysname{}-to-Greedy compilation-time ratio averages 1.03 across the device-graph and width means and spans 0.84 to 1.12 on an AMD Ryzen 5 5600X (Fig.~\ref{fig:eval-scaling-impact}c). Scoring an unresolved operation $g$ at a fixed layout takes $O\!\left(\sum_{r\in\mathcal{R}(g)}|I_r|\right)$ time, a constant for the dual-rail ISA. Appendix~\ref{app:router-complexity} gives the router's time and space bounds, which are the same for both compilers.

\subsection{Ablations and Sensitivity}\label{sec:eval-ablations}

\begin{figure}[t]
  \centering
  \includegraphics[width=\columnwidth]{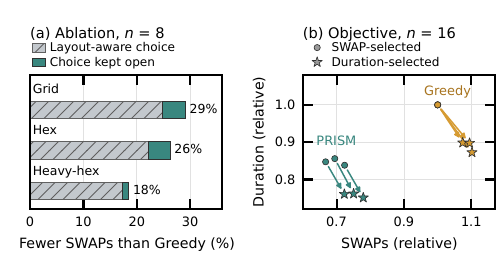}
  \caption{Panel (a) shows routing SWAP reduction relative to Greedy at eight logical qubits. Gray bars (Layout-aware choice) show an ablated \sysname{} that commits gates on entering the front layer, and teal bars (Choice kept open) show \sysname{}'s further reduction. Panel (b) plots routing SWAPs against duration at 16 logical qubits relative to Greedy, with one arrow per device graph from SWAP-selected to duration-selected routes.}
  \label{fig:eval-ablations}
\end{figure}

On the 16 MQT workloads, an ablated \sysname{} commits each gate when it first enters the front layer, retaining layout-aware choice but preventing further layout changes before commitment (Fig.~\ref{fig:eval-ablations}a). On entry it commits an executable realization if one exists and otherwise the nearest one under that layout (Eq.~\eqref{eq:greedy-policy}), and it scores lookahead gates by Eq.~\eqref{eq:prism-routing-cost} as \sysname{} does. Over 20 seeds, the layout-aware choice alone reduces Greedy's total routing SWAP count by 24.8\% on the grid, 22.2\% on the hexagonal lattice, and 17.3\% on heavy-hex, and the choice kept open reduces the ablated count by a further 5.8\%, 5.2\%, and 1.4\%. The ablated compiler averages more routing SWAPs than Greedy only for qaoa on the hexagonal lattice; \sysname{} does not show this increase.

\sysname{} also breaks ties among executable realizations by usage count, which Greedy never needs. Disabling the counts leaves every per-run SWAP count of the 16 MQT workloads at eight logical qubits on 20 seeds unchanged, and \sysname{}'s depth reductions on the grid and hexagonal lattice move from 11.1\% and 10.4\% to 9.9\% and 9.9\%, so commitment timing alone accounts for the SWAP difference.

On 25- and 36-site grids at eight logical qubits, which add spare sites, Greedy's total routing SWAP count stays within 0.3\% of its 20-site value, and \sysname{} reduces it by 28.8\% and 28.5\%.

On the 16 MQT workloads at 16 logical qubits, selecting completed routes by depth or critical-path duration instead of SWAP count shifts both compilers alike, trading 6--10\% more SWAPs for 9--13\% shorter schedules (Fig.~\ref{fig:eval-ablations}b). \sysname{}'s advantage stays at 27.6--34.4\% fewer SWAPs and 12.5--16.4\% shorter schedules under every objective. The SWAP search still scores distance, so only the final route choice changes.

Selecting Greedy's realizations from the sites where each pair is prepared, instead of from each call's entry layout, changes 11\% of its selections in a sample of 120 runs. This alternative changes Greedy's routing SWAPs by $-2.7\%$ to $+1.2\%$ per device graph and lengthens its critical-path duration by 1.3\% to 2.4\%. Logical-SABRE's routing SWAP count depends on its site matching. At eight logical qubits, using eleven alternative maximum matchings on each graph changes Logical-SABRE's routing SWAP count relative to its original matching, yielding 24\% fewer to 9\% more SWAPs on the grid and 0.5\% to 36\% fewer on heavy-hex. Four of the heavy-hex matchings leave a disconnected logical coupling graph on which the circuit cannot be placed. Logical-SABRE's fallback to the aligned circuit (Fig.~\ref{fig:logical-swap-candidates}a) lowers 50\% of its logical SWAPs on the grid and every logical SWAP on the hexagonal lattice and heavy-hex. The SWAPs that its physical routing pass adds make up 35\%, 53\%, and 48\% of its routing SWAPs on the grid, hexagonal lattice, and heavy-hex.

\subsection{Shorter Routes Reduce Shot Overhead}\label{sec:eval-device}

\begin{figure}[t]
  \centering
  \includegraphics[width=\columnwidth]{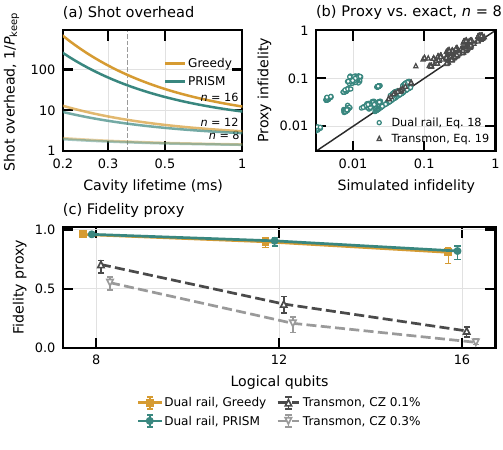}
  \caption{Hardware projections use the 16 MQT workloads (five seeds) on the grid, hexagonal lattice, and heavy-hex. Panel (a) plots shot overhead $P_{\mathrm{keep}}^{-1}$ (Eq.~\eqref{eq:hardware-loss-model}) against cavity lifetime, geometrically averaged over workloads at 8, 12, and 16 logical qubits. The dashed line marks the measured device. Panel (b) compares proxy infidelity (Eqs.~\eqref{eq:hardware-residual-model} and~\eqref{eq:transmon-fidelity}) with exact simulation at eight logical qubits, with transmons on the grid and hexagonal lattice only. Panel (c) shows median fidelity proxy and interquartile range. Dual-rail fidelity is over retained shots, and the transmon retains every shot.}
  \label{fig:eval-physical-proxies}
\end{figure}

Terminal erasure checks discard shots with photon loss, so the probability of loss along a route determines its shot overhead, the expected number of attempts per retained shot. Residual dephasing and bit flips are not flagged by the erasure checks, so affected shots remain in the retained sample and reduce its fidelity. We project both costs for the scaling routes at 8, 12, and 16 logical qubits (Table~\ref{tab:device-models}).

\begin{table}[t]
  \centering
  \caption{Device models~\cite{mehta_bias-preserving_2025,abughanem_ibm_2025,google_below_threshold_2025}. Dual-rail CZ is the physical $\mathrm{CR}_z$ (SWS) gate, with values for its control/target pair (C/T), and dual-rail 1Q is $R_{\mathrm{SWAP}}$, averaged over the $\sqrt{X}$ gates measured on the control and target qubits. Dual-rail $T_1$ is the reciprocal of the mean decay rate of the four cavities, and $T_\phi$ is the modeled coherence decay time of a rail pair (Section~\ref{sec:eval-device}). Erasure is per rail pair touched, and residual is per operation. Transmon residuals are $r_g$ in Eq.~\eqref{eq:transmon-fidelity}.}
  \label{tab:device-models}
  \footnotesize
  \setlength{\tabcolsep}{4pt}
  \begin{tabular}{@{}lcccc@{}}
    \toprule
    & \multicolumn{2}{c}{Dual rail} & \multicolumn{2}{c}{Transmon} \\
    \midrule
    $T_1$ & \multicolumn{2}{c}{$357\,\mu$s} & \multicolumn{2}{c}{$137\,\mu$s} \\
    $T_\phi$ & \multicolumn{2}{c}{$2.2$\,ms} & \multicolumn{2}{c}{$132\,\mu$s ($T_2 = 89\,\mu$s)} \\
    \devicerules
    & \makebox[1.2cm]{CZ} & \makebox[1.2cm]{1Q} & \makebox[1.2cm]{CZ} & \makebox[1.2cm]{1Q} \\
    Duration (ns) & 500 & 200 & 42 & 25 \\
    Erasure ($10^{-4}$) & 40 / 9.6 & 7.2 & -- & -- \\
    Residual ($10^{-4}$) & 3.9 / 1.1 & 0.99 & 10 or 30 & 3 \\
    \bottomrule
  \end{tabular}
\end{table}

The dual-rail model in Table~\ref{tab:device-models} takes its parameters from measurements of Mehta et al.'s cavity device~\cite{mehta_bias-preserving_2025}, and the transmon model uses typical published parameters~\cite{abughanem_ibm_2025,google_below_threshold_2025} with two tested CZ errors. Mehta et al. estimate the phase-flip probability of an idle dual-rail qubit as $t/T_{\mathrm{echo}}$ and report echo times $T_{\mathrm{echo}}$ of 4.0 and 4.8\,ms for their two dual-rail qubits. We set $\gamma$ to the mean of the two rates $1/T_{\mathrm{echo}}$, $\gamma \approx (4.4\,\mathrm{ms})^{-1}$, so an idle rail pair suffers a phase flip with probability $(1-e^{-2\gamma t})/2 \approx \gamma t$ and its coherence decays as $e^{-2\gamma t}$, with decay time $T_\phi = 1/(2\gamma) = 2.2$\,ms. A cavity rail dephases far more slowly than it loses its photon ($T_\phi \gg T_1$), so most of its errors are detected, whereas a transmon has $T_\phi \approx T_1$. Both models assume ideal preparation, readout, and erasure checks. For every $R_{\mathrm{SWAP}}$ angle and routing SWAP between pairs, our model uses the values measured for the $\sqrt{X}$ beamsplitter gate, since the two-qubit device characterizes neither a full SWAP nor one between pairs.

Each dual-rail qubit holds one excitation for the whole circuit, and its rail pair loses that excitation at rate $1/T_1$ while it idles. Each operation also erases each pair it touches with a measured probability. The probability that a shot passes the terminal erasure checks is
\begin{equation}
  P_{\mathrm{keep}}
    = \exp\!\left(-\frac{T_{\mathrm{idle}}}{T_1}\right)\prod_{g}(1-e_g)^{K_g},
  \label{eq:hardware-loss-model}
\end{equation}
where $g$ ranges over three roles (the control pair of a $\mathrm{CR}_z$, its target pair, and a pair in an $R_{\mathrm{SWAP}}$ or routing SWAP), $e_g$ is the erasure probability of role $g$, and $K_g$ counts the rail pairs touched in that role, so each $\mathrm{CR}_z$ adds one to the control count and one to the target count. The idle time $T_{\mathrm{idle}}$ sums, over the rail pairs that hold a logical qubit, the time within the scheduled duration $[0,D]$ during which neither rail of the pair is in an operation. The model treats every pair as prepared at time zero by an instantaneous, error-free preparation, so $T_{\mathrm{idle}}$ and $K_g$ include the time and routing SWAPs before a pair's preparation. The shot overhead is $P_{\mathrm{keep}}^{-1}$~\cite{griffiths_limits_2026}. For the retained shots, following success-probability models~\cite{baker_exploiting_2021}, the product of the idle coherence factor and the gate fidelities gives the fidelity proxy
\begin{equation}
  F_{\mathrm{cond}}
    = e^{-\gamma T_{\mathrm{idle}}}
      \prod_{g\in\{\mathrm{C},\mathrm{T},R_{\mathrm{SWAP}}\}}
      (1-\epsilon_g)^{N_g}
  \label{eq:hardware-residual-model}
\end{equation}
for operation counts $N_g$ with residual errors $\epsilon_g$, where $N_{\mathrm{C}}$ and $N_{\mathrm{T}}$ both equal the $\mathrm{CR}_z$ count and $N_{R_{\mathrm{SWAP}}}$ counts each logical $R_X$ realization and full-angle routing SWAP once, however many pairs it touches. For a SWAP between two pairs, we exclude its duration from both pairs' idle time, apply the SWAP erasure probability to each of the two rail pairs in Eq.~\eqref{eq:hardware-loss-model}, and use one factor $1-\epsilon_{R_{\mathrm{SWAP}}}$, although $\epsilon_{R_{\mathrm{SWAP}}}$ was measured on a single pair. The factor $e^{-\gamma T_{\mathrm{idle}}}$ is a first-order approximation to, and a lower bound on, the probability that no pair suffers a logical phase flip while it idles. We aggregate each quantity as the geometric mean of the per-workload \sysname{}-to-Greedy ratios.

At 16 logical qubits, \sysname{} shortens the route of every workload, reducing total duration by 15.3\%, 14.4\%, and 16.2\% on the grid, hexagonal lattice, and heavy-hex (11.5\%, 13.7\%, and 15.0\% with equal workload weights). Because the probability of keeping a shot decays exponentially with idle time (Eq.~\eqref{eq:hardware-loss-model}), the shot overhead falls by more than the duration. It falls on every workload but one, by 41.8\%, 42.9\%, and 48.1\% as geometric means over workloads (44.3\% over workloads and all three device graphs, Fig.~\ref{fig:eval-physical-proxies}a) and by 18.2\%, 26.6\%, and 28.7\% for the median workload. The proxy infidelity $1-F_{\mathrm{cond}}$ also falls on every workload but one, by 7.4\%, 9.1\%, and 10.0\% as geometric means, a smaller gain because erasure conversion~\cite{wu_erasure_2022,kubica_erasure_2022} leaves few residual errors for a shorter route to remove.

For an unencoded transmon, which retains every shot, the same construction gives the fidelity proxy
\begin{equation}
  F_{\mathrm{T}} = \exp\!\left[-T_{\mathrm{idle}}^{\mathrm{T}}\left(\frac{1}{4T_1}+\frac{1}{2T_2}\right)\right]\prod_{g}\left(1-\frac{d_g+1}{d_g}\,r_g\right)^{N_g},
  \label{eq:transmon-fidelity}
\end{equation}
where $T_{\mathrm{idle}}^{\mathrm{T}}$ is the summed idle time of the active transmons, $r_g$ is the average error of gate type $g$, and $d_g = 2^k$ for a $k$-qubit gate. The gate factor converts average to process fidelity~\cite{nielsen_simple_2002}, and the idle factor follows by the same conversion from the average idle fidelity under relaxation and dephasing~\cite{bao_fluxonium_2022}.

We check Eqs.~\eqref{eq:hardware-loss-model}--\eqref{eq:transmon-fidelity} by simulating the routed circuits under the same channels. At eight logical qubits we simulate exactly, with density matrices in Qiskit, all 480 dual-rail circuits on the three device graphs and 160 transmon circuits on the grid and hexagonal lattice. At 16 logical qubits we sample 200 quantum trajectories for each of the 320 dual-rail circuits on the grid and hexagonal lattice. The simulated acceptance equals Eq.~\eqref{eq:hardware-loss-model} on every circuit. The fidelity proxies treat every residual error as fatal, although many leave the output state unchanged, so they overstate infidelity by a median factor of 1.6 at eight logical qubits and 1.8 at 16 for dual rail and 1.2 for the transmon (Fig.~\ref{fig:eval-physical-proxies}b).

Under a global-depolarizing assumption that counts both discarded shots and residual errors, dual rail needs less sampling time than an unencoded transmon by 16 logical qubits with either compiler, and \sysname{} lowers the ratio at every width (Fig.~\ref{fig:eval-physical-proxies}c). The transmon uses one qubit per logical qubit. Qiskit compiles each circuit onto the same device graphs at optimization level 3 into the basis $\{\mathrm{CZ},\sqrt{X},X,R_z\}$ with free $R_z$, and we keep the candidate with the highest estimated success probability among 80, from 40 transpiler seeds on the circuit before and after our logical stage. From 8 to 16 logical qubits the transmon's fidelity proxy at 0.1\% CZ error falls from 0.70 to 0.14, while that of the retained dual-rail shots with \sysname{} stays between 0.96 and 0.82. To compare sampling time, we assume that the residual noise acts as global depolarizing noise, which local noise approaches in scrambling circuits~\cite{dalzell_whitenoise_2024}. A fidelity $F$ then scales each observable's signal by $F$, so a fixed precision takes $1/F^2$ retained shots~\cite{vovrosh_depolarizing_2021,takagi_mitigation_2022,griffiths_limits_2026}, and sampling time is proportional to the critical-path duration divided by $P_{\mathrm{keep}}F^2$ for dual rail and by $F^2$ for the transmon. Because dual-rail residual errors are mostly dephasing, which leaves $Z$-basis observables unattenuated, this assumption overstates the dual-rail cost. On the grid and hexagonal lattice at 8, 12, and 16 logical qubits, dual rail needs 2.5, 1.2, and 0.34 times the transmon's sampling time at 0.1\% CZ error with \sysname{} and 2.9, 1.8, and 0.72 times with Greedy, as geometric means over paired circuits. If the transmon instead ran two copies on the qubits that dual rail spends on second rails, every ratio would at most double, so at 16 logical qubits \sysname{} would still need at most 0.67 times the transmon's time and Greedy at most 1.44 times. These sampling-time estimates omit preparation, readout, reset, state-dependent coupler loss, erasure-check overhead, correlated errors, and decoding.

Of the varied hardware parameters, cavity lifetime matters most (Fig.~\ref{fig:eval-physical-proxies}a). At 16 logical qubits, the median workload needs 10.5 attempts per retained shot with \sysname{} and 12.6 with Greedy, and twice the cavity lifetime lowers these to 4.7 and 5.2, against 8.1 and 9.6 for a beamsplitter twice as fast and 9.3 and 11.7 for half the beamsplitter erasure probability. An erasure check that detects only a fraction $\eta$ of erasures leaves the rest in the retained shots. The median fidelity proxy falls from 0.82 to 0.81 at Mehta et al.'s measured $\eta = 99.6\%$~\cite{mehta_bias-preserving_2025} and to 0.75 at the $\eta = 96.3\%$ of de Graaf et al.'s check~\cite{degraaf_midcircuit_2025} with \sysname{}, and from 0.81 to 0.80 and 0.74 with Greedy. Doubling the residual gate errors lowers the two medians to 0.76 and 0.75. \sysname{} keeps its advantage under every variation, but at the measured parameters the median workload keeps only 10\% of its shots at 16 logical qubits, motivating an outer code that can correct losses instead of discarding every affected shot.

\subsection{Shorter Routes Lower Decoded Logical Error}\label{sec:outer-code}

Under an outer code, the decoder learns of an erasure only when a check or readout flags it. Checks inside the circuit flag erasures sooner, but each check forces the two rails of its pair to be adjacent. We compile a three-round $[[7,1,3]]$ Steane memory~\cite{steane_multiple_1996} using the qLDPC package's default extraction circuit~\cite{perlin_qldpc_2023} on the 48-site hexagonal lattice, where seven data and six ancilla qubits require 26 rails or 13 transmons. The circuit measures X checks with CX and Z checks with CZ from ancillas prepared in $|+\rangle$, and the X and Z memories, which differ in preparation and readout, are compiled separately. After lowering (Appendix~\ref{app:steane-noise}), each memory has 72 $\mathrm{CR}_z$ gates and 78 $R_X(\pm\pi/2)$ gates, each an $R_{\mathrm{SWAP}}$ that requires a qubit's rails to be adjacent.

With flags at readout only, the circuit has no intermediate erasure checks and each pair's two-rail readout flags its erasures. With compiled checks, each data pair also receives a joint erasure check after its last two-qubit gate of rounds one and two, and both compilers route the 14 checks like other two-rail operations. Ancilla pairs are read in every round. The checks follow de Graaf et al.'s measured check~\cite{degraaf_midcircuit_2025} on a different device, and the other faults follow Table~\ref{tab:device-models}. We model an erasure as skip-gate leakage~\cite{pavlovich_erasure_2026}. A lost pair is empty, every gate it takes part in does not happen until a check or readout flags it, and each flag resets the pair to a completely mixed state. Stim's tableau simulator~\cite{gidney_stim_2021} samples every fault at its scheduled location under the channels of Appendix~\ref{app:steane-noise}. A belief-propagation and ordered-statistics (BP+OSD) decoder~\cite{panteleev_degenerate_2021,roffe_decoding_2020} receives erasure information only through the flags, and for each flag it includes the errors that a loss at any of the pair's locations since its last observation could have caused.

Greedy and \sysname{} use twenty paired routing seeds, the search budget of Section~\ref{sec:eval-setup}, and no final rescheduling. Each compiler completes one route per initial layout, 43 in all, and keeps the one most likely to have no erasure or $Z$ fault. It minimizes the sum of $-\ln(1-e)-\ln(1-p_Z)$ over the scheduled fault locations, where $e$ and $p_Z$ are the erasure and $Z$ fault probabilities of a location, with separate $\mathrm{CR}_z$ control and target rates and each pair's idle time from preparation to readout, while candidate SWAPs are still scored by distance and final trials by SWAP count. The transmon baseline is compiled as in Section~\ref{sec:eval-device} on the same graph with the same 16 refinement iterations and 40 final trials, is selected among its 80 candidate routes by the analogous no-fault probability, and has its thermal channel Pauli twirled~\cite{geller_efficient_2013} for Stim. Its comparison with dual rail depends on the hardware models, the transmon's instantaneous readout and reset, and the fixed extraction circuit.

\begin{table}[t]
  \centering
  \caption{Decoded logical failure probability $P_L$ of the three-round Steane memory on the hexagonal lattice under skip-gate leakage, averaged over twenty routes with 100,000 shots each, or 200,000 for the transmon rows. Brackets are 95\% paired seed-bootstrap intervals. Reductions compare means.}
  \label{tab:outer-code}
  \footnotesize
  \setlength{\tabcolsep}{4pt}
  \begin{tabular}{@{}lrr@{}}
    \toprule
    Implementation & X memory $P_L$ & Z memory $P_L$ \\
    \midrule
    \emph{Flags at readout only} & & \\
    \quad Dual rail, Greedy & 0.7645\% & 0.7165\% \\
    \quad Dual rail, \sysname{} & 0.6648\% & 0.5760\% \\
    \quad Relative reduction & 13.0\% [2.1, 22.5] & 19.6\% [12.3, 26.3] \\
    \emph{Compiled erasure checks} & & \\
    \quad Dual rail, Greedy & 1.2070\% & 1.2607\% \\
    \quad Dual rail, \sysname{} & 0.8728\% & 0.9756\% \\
    \quad Relative reduction & 27.7\% [21.6, 32.8] & 22.6\% [13.8, 30.5] \\
    \midrule
    Transmon, 0.1\% CZ error & 2.1724\% & 2.0526\% \\
    Transmon, 0.3\% CZ error & 5.6250\% & 5.2801\% \\
    \bottomrule
  \end{tabular}
\end{table}

With flags at readout only, \sysname{} lowers $P_L$ by 13.0\% in the X memory and 19.6\% in the Z memory (Table~\ref{tab:outer-code}), on 14 and 18 of 20 seeds, with the lower confidence bound for the X-memory reduction near zero. Greedy's routes have more loss locations, so more of its shots lose two pairs, and such shots cause most failures. With checks, \sysname{} lowers $P_L$ by 27.7\% and 22.6\%, on 19 and 17 seeds. Its checked routes use 42\% and 35\% fewer routing SWAPs than Greedy's and shorten the circuits from 20.8 to 17.3~$\mu$s and from 20.9 to 17.4~$\mu$s. The checks add one to four SWAPs to \sysname{}'s routes and 3 to 14 to Greedy's, so \sysname{} retains its advantage despite the checks' adjacent-rail requirement. At these check parameters, checks raise both compilers' $P_L$, Greedy's to 1.6 and 1.8 times and \sysname{}'s to 1.3 and 1.7 times their readout-only values.

\section{Discussion}\label{sec:discussion}

\subsection{Interpreting the Results}\label{sec:interpreting-results}

Greedy and \sysname{} share the boundary IR, the router, and the search budget, so their performance difference reflects the effect of commitment timing. Logical-SABRE also routes on a fixed logical coupling graph over a fixed site matching, so its gap mixes commitment timing with that design.

Choosing realizations before routing and choosing them during routing have the same global optimum, because every complete route fixes a realization assignment. Finding that optimum requires optimizing over $m^{N_g}$ assignments of $m$ realizations to $N_g$ gates and solving a qubit-routing problem that is NP-hard even on a line~\cite{ito_algorithmic_2023}. \sysname{} uses a SABRE-based heuristic, which does not guarantee an optimal route~\cite{li_qrob_2026}.

The available choices depend on how long a pair's rails can remain apart. Logical $X$, Hadamard, and general $R_X$ rotations need the two rails to be adjacent because only $R_Z$ has an admissible product completion (Eq.~\eqref{eq:loss-compatible-complement}). Intermediate erasure checks also require adjacent rails (Section~\ref{sec:outer-code}). Between these operations the rails may separate, and the longer they stay apart, the more realization choices routing can use. While its rails are apart, a qubit can take part only in gates that preserve its logical $Z$ value, because each operation touches at most one of its rails. Teleportation and inter-module communication~\cite{devulapalli_quantum_2024,wu_autocomm_2022}, measurement and feedforward~\cite{baumer_dynamic_2024,baumer_qft_2024}, ancilla-mediated embeddings~\cite{berthusen_2dlocal_2025}, and microwave links between modules~\cite{teoh_links_2025} lie outside this model.

\subsection{Limitations and Future Work}\label{sec:limitations}

A realization set holds only circuits that act on fixed wires, leave the same wire mapping and logical frame, and embed without SWAPs. For the logical SWAP inserted by Logical-SABRE, variant (c) leaves an $X_L$ on each qubit, which a compiler that tracked Pauli frames~\cite{chamberland_frame_2017} could absorb, and its other variants need two inter-pair edges at one layout. These restrictions limit the retained realizations, not the programs \sysname{} can compile, because every logical $\mathrm{CR}_z$ has a single-interaction realization that the router can make executable on any connected device graph. The definitions in Sections~\ref{sec:lowering-partial-spec} and~\ref{sec:logical-availability} can be used with other encodings once their admissible realization circuits and interaction patterns are specified. Such circuits may need routing between their own interactions. A logical graph weighted by routing cost would then generalize the availability graph.

The router keeps unresolved $\mathrm{CR}_z$ gates that share a logical qubit in program order, because each depends on all four rails of its operands, although their diagonal realizations commute. An executable gate can therefore wait behind one that still needs routing. The order constrains routing but not the emitted circuit, in which each committed gate depends only on its own two rails, so gates committed to disjoint rails of the shared qubit run in the same layer. Jointly selecting realizations for commuting gates could relax this ordering constraint.

Each realization also fixes which operand's rail is the SWS control. Because $\mathrm{CP}(\theta)$ is symmetric (Eq.~\eqref{eq:cp-crz-convention}), exchanging the roles costs only software $R_Z$ rotations and would double each $\mathrm{CR}_z$ realization set. We have not evaluated this choice, which could matter under an outer code because losing the control also dephases the target (Appendix~\ref{app:steane-noise}).

The boundary IR does not depend on SABRE. Lookahead, exact, and time-aware routers such as t$|$ket$\rangle$'s~\cite{cowtan_routing_2019}, OLSQ~\cite{tan_olsq_2020}, and TOQM~\cite{zhang_toqm_2021} could use the boundary IR by testing which realizations embed under each layout and keeping alternatives until commitment.

\section{Related Work}\label{sec:related}

Synthesis supplies circuits for realization sets. Logical Clifford synthesis finds low-cost physical Clifford circuits~\cite{rengaswamy_logical_2020}, hardware-tailored synthesis uses freedom outside the code space to satisfy connectivity~\cite{kuehnke_hardware_2025}, stabilizer equivalence removes inaccessible interactions~\cite{weinberg_stabilizers_2026}, and code automorphisms~\cite{grassl_automorphisms_2013,sayginel_automorphisms_2024} and chain maps~\cite{benhemou_chainmaps_2026} give circuits for a logical action.

Iceberg co-compilation jointly searches the algorithmic circuit and flexible fault-tolerant gadgets~\cite{jin_iceberg_2025}, and multi-qubit Iceberg compilation assigns program qubits to logical positions~\cite{mazzetti_logical_2026}. Surface-code compilers represent logical operations through patches, measurement regions, and routing paths~\cite{litinski_game_2019,molavi_dependency_2025,watkins_high-performance_2024}, interleaved logical qubits derive connectivity from their physical constituents~\cite{viszlai_interleaved_2025}, and quantum low-density parity-check (qLDPC) compilation combines code symmetries with hardware routing~\cite{liu_toric_2026}. Each targets a particular code and architecture, and none of them retains an encoded gate's alternative realizations through physical routing. Marol's device state machine~\cite{molavi_marol_2026} tracks where a fixed instruction can execute but does not retain alternative realizations.

The coupling between realization choice and layout is the phase-ordering problem of classical compilers. Instruction selection chooses among machine implementations of one IR operation~\cite{hjort_blindell_instruction_2016,koes_near-optimal_2008}, register allocation and scheduling set each choice's cost, and committing either phase first can degrade the other~\cite{vegdahl_phase_1982}. Classical compilers exchange partial results between phases~\cite{bradlee_integrating_1991,pinter_register_1993,ambrosch_dependence_1994,norris_experimental_1995}, search both phases jointly~\cite{castaneda_lozano_combinatorial_2019}, or delay binding until later phases supply the required context~\cite{wulf_delayed_1982}.

Quantum mappers that exploit gate equivalence choose among implementations on the same physical qubits. They apply commutation rules~\cite{itoko_optimization_2020}, reorder Trotter terms~\cite{lao_2qan_2022}, resynthesize blocks under a free output permutation~\cite{liu_permutation_2023}, or estimate SWAP costs from later savings~\cite{liu_not_2022}. Orchestrated Trios routes a Toffoli's operands before selecting its decomposition~\cite{duckering_trios_2021}, QContext selects decomposition variants after routing~\cite{liu2023qcontext}, and lazy synthesis interleaves SWAP insertion with synthesis~\cite{martiel_lazy_synthesis_2022}. MIRAGE is the closest prior approach, because its SABRE-style loop chooses between a two-qubit gate and a mirrored form that absorbs a SWAP~\cite{mckinney_mirage_2024}. An encoded gate's realizations instead use different wires and so need different device edges.

\section{Conclusion}

Routing changes which realization of an encoded gate can execute without SWAPs, so \sysname{} keeps each gate's realizations open until the router can execute one. For dual-rail cavity qubits, this cuts Greedy's routing SWAPs by 20.4--30.8\% and physical depth by 8.8--10.1\% across 25 workloads on three device graphs. On modeled cavity hardware, the shorter routes need 44.3\% fewer attempts per retained shot at 16 logical qubits and lower decoded error in a three-round Steane memory.

\section*{Code Availability}
\sysname{} is open source under the Apache-2.0 license at \url{https://github.com/Yale-QCS/prism} and installs as the Qiskit package \texttt{prism-qc}. The repository also includes the code, scripts, and data that reproduce every figure and number in this paper.

\section*{Acknowledgments}
The authors thank Joe Aulicino, Raj Pabari, Nishaad Khedkar, and Chao Zhou for helpful discussions. This material is based upon work supported in part by the U.S. Department of Energy, Office of Science, National Quantum Information Science Research Center, Co-design Center for Quantum Advantage (C2QA) under Contract No. DE-SC0012704, and in part by Air Force Office of Scientific Research MURI (FA9550-26-1-B036). Y.D. acknowledges partial support by the National Science Foundation (under awards CCF-2312754 and CCF-2338063), by Quantum CT (under NSF Engines award ITE-2302908), by Boehringer Ingelheim, by NSF NQVL-ERASE (under award OSI-2435244), and by NSF QLCI Practiqal (under award OSI-2552449). External interest disclosure: Y.D. is a consultant and equity holder of Toric Labs and D-Wave Quantum. R.J.S. is a consultant and equity holder of D-Wave Quantum.

\bibliographystyle{IEEEtran}
\bibliography{refs}

\appendices
\section{Weyl-Chamber Coordinates of Dual-Rail Completions}\label{app:weyl-derivation}
The surfaces and lines in Fig.~\ref{fig:weyl-sweep} follow from the canonical form of parity-preserving gates~\cite{brod2011extending}. Write the nonlocal part of a two-rail unitary as $U_{\mathrm{nl}}=\exp[i(c_1X\otimes X+c_2Y\otimes Y+c_3Z\otimes Z)/2]$, so that $(c_1,c_2,c_3)$ are its Weyl-chamber coordinates, and for the blocks of Eq.~\eqref{eq:GAB} let $\cos\phi_B=|B_{11}|$ and $\cos\phi_A=|A_{11}|$. Before the coordinates are folded into the Weyl chamber, $c_1=\phi_B+\phi_A$, $c_2=\phi_B-\phi_A$, and $c_3$ is set by the relative phase of $\det A$ and $\det B$. Fixing the logical gate fixes $B$ and the unfolded sum $c_1+c_2=2\phi_B$, while varying $A$ sweeps the folded surface. The surface reaches the local equivalence class at the origin only when $\phi_B\in\{0,\pi/2\}$, which means that $B$ is diagonal or anti-diagonal and gives the family of Eq.~\eqref{eq:delocal-group}.

An admissible completion has the diagonal $A$ of Eq.~\eqref{eq:loss-compatible-complement}, so $\phi_A=0$. Since $\det A=e^{2i\alpha}$ does not depend on $\beta$, the phase $\alpha$ alone sweeps $c_3$ along the orange line $c_1=c_2=\phi_B$. For logical $X$, $\phi_B=\pi/2$ and this line is the parametric-SWAP edge from $\mathrm{iSWAP}$ to $\mathrm{SWAP}$.

\section{Device Graph Dimensions}\label{app:device-graphs}
Table~\ref{tab:device-graphs} lists the device graphs at every evaluated width $n$. All three come from Qiskit's coupling-map generators. The grid has $R=\lceil\sqrt{2n+2}\,\rceil$ rows and $C=\lceil(2n+2)/R\rceil$ columns. The hexagonal lattice has $k\times k$ hexagons with $k=\lceil\sqrt{n+1}\,\rceil$, which gives $2(k+1)^2-2$ sites. The heavy-hex graph has the smallest odd distance $d\geq 3$ whose $(5d^2-2d-1)/2$ sites hold the workload. Under the $k\times k$ rule the hexagonal lattice has more spare sites than the grid at every width, and at ten of the 13 widths the next smaller $k\times k$ lattice would also hold the workload. MQT Bench's VBE adder on $m$-bit operands has $3m+1$ qubits, so it exists at 16 and 28 logical qubits. At the other widths it has 7, 10, 19, 22, and 31 logical qubits and uses the graphs for $n=8$, 12, 20, 24, and 32. The larger grids of Section~\ref{sec:eval-ablations} are $5\times5$ and $6\times6$.

\begin{table}[t]
  \centering
  \caption{Site counts appear in parentheses for each width $n$. Heavy-hex is not evaluated at $n=24$, 28, and 32.}
  \label{tab:device-graphs}
  \footnotesize
  \setlength{\tabcolsep}{4pt}
  \begin{tabular}{@{}rrlll@{}}
    \toprule
    $n$ & $2n+2$ & Grid $R\times C$ & Hexagonal $k$ & Heavy-hex $d$ \\
    \midrule
     8 & 18 & $5\times4$ (20) & 3 (30) & 3 (19) \\
    11 & 24 & $5\times5$ (25) & 4 (48) & 5 (57) \\
    12 & 26 & $6\times5$ (30) & 4 (48) & 5 (57) \\
    13 & 28 & $6\times5$ (30) & 4 (48) & 5 (57) \\
    15 & 32 & $6\times6$ (36) & 4 (48) & 5 (57) \\
    16 & 34 & $6\times6$ (36) & 5 (70) & 5 (57) \\
    17 & 36 & $6\times6$ (36) & 5 (70) & 5 (57) \\
    19 & 40 & $7\times6$ (42) & 5 (70) & 5 (57) \\
    20 & 42 & $7\times6$ (42) & 5 (70) & 5 (57) \\
    24 & 50 & $8\times7$ (56) & 5 (70) & \\
    25 & 52 & $8\times7$ (56) & 6 (96) & 5 (57) \\
    28 & 58 & $8\times8$ (64) & 6 (96) & \\
    32 & 66 & $9\times8$ (72) & 6 (96) & \\
    \bottomrule
  \end{tabular}
\end{table}

\begin{table}[t]
  \centering
  \caption{Of the 25 workloads in Fig.~\ref{fig:eval-core-benchmarks}, those from MQT Bench 2.2.2 are algorithm-level benchmarks with $n=8$, except vbe with $n=7$. The 16 MQT workloads used for scaling appear above the horizontal rule. The QASMBench medium-scale files end in their width $n$, which the labels omit.}
  \label{tab:workloads}
  \footnotesize
  \setlength{\tabcolsep}{4pt}
  \begin{tabular}{@{}ll@{}}
    \toprule
    Label & MQT Bench benchmark \\
    \midrule
    dj & \texttt{dj} \\
    qaoa & \texttt{qaoa} \\
    draper & \texttt{draper\_qft\_adder} \\
    mod-add & \texttt{modular\_adder} \\
    qft & \texttt{qft} \\
    qpe-ex & \texttt{qpeexact} \\
    qpe-in & \texttt{qpeinexact} \\
    qft-ent & \texttt{qftentangled} \\
    copula & \texttt{bmw\_quark\_copula} \\
    hhl & \texttt{hhl} \\
    ae & \texttt{ae} \\
    cdkm & \texttt{cdkm\_ripple\_carry\_adder} \\
    vbe & \texttt{vbe\_ripple\_carry\_adder} \\
    vqe-2local & \texttt{vqe\_two\_local} \\
    rg-mult & \texttt{rg\_qft\_multiplier} \\
    mult & \texttt{multiplier} \\
    \workloadrule
    randomcircuit & \texttt{randomcircuit} \\
    \midrule
    Label & QASMBench file \\
    \midrule
    sat & \texttt{sat\_n11} \\
    seca & \texttt{seca\_n11} \\
    multiply & \texttt{multiply\_n13} \\
    qf21 & \texttt{qf21\_n15} \\
    qec9xz & \texttt{qec9xz\_n17} \\
    bv & \texttt{bv\_n19} \\
    qram & \texttt{qram\_n20} \\
    knn & \texttt{knn\_n25} \\
    \bottomrule
  \end{tabular}
\end{table}

\section{Router Complexity}\label{app:router-complexity}
For $L$ initial layouts with $\rho$ forward--backward refinements and $\tau$ final trials each, $N$ lowered operations, at most $S$ inserted SWAPs per pass, and $n$ logical qubits on a device of $n_{\mathrm{phys}}$ sites and maximum degree $\Delta$, both compilers run the same router in time
\begin{equation}
  O\bigl(L(2\rho+\tau)\bigl(N+S(n_{\mathrm{phys}}+n\Delta\log(n\Delta))\bigr)+L(2\rho+1)\,n_{\mathrm{phys}}^3\bigr)
  \label{eq:router-complexity}
\end{equation}
and space $O(n_{\mathrm{phys}}^2+N)$, plus $O(N_{\mathrm{out}}^2)$ time to schedule the $N_{\mathrm{out}}$ gates of each scheduled route. The $n\Delta\log(n\Delta)$ term is the per-step sort of candidate SWAPs by physical endpoints, and the $n_{\mathrm{phys}}$ term is a per-step scan of all device sites that finds the rails of unresolved front-layer gates and decays the usage counts. For each candidate SWAP, the router rescores only the front-layer gates on its two sites and at most 20 lookahead gates, so scoring takes constant time. The $n_{\mathrm{phys}}^3$ term recomputes the distances between device sites, by a bit-parallel breadth-first search from every site, at the start of each of the $2\rho+1$ router calls per layout. The $O(n_{\mathrm{phys}}^2+N)$ space bound accounts for the distance matrix and routing state.

\section{Steane Memory Noise Model}\label{app:steane-noise}
\textit{Circuit.} Lowering writes $H$ as $R_Z(\pi/2)R_X(\pi/2)R_Z(\pi/2)$, CX as CZ conjugated by $H$ on the target, and CZ as $\mathrm{CR}_z(\pi)$ with an $R_Z$ correction, and the transmon baseline uses the same rules over CZ, $\sqrt{X}$, $X$, and $R_z$. The logical stage absorbs the Hadamard gate of each $|+\rangle$ preparation into the checks, so the ancilla pair is the $\mathrm{CR}_z$ control in the Z checks of the X memory and in the X checks of the Z memory.

\textit{Faults.} Every fault location belongs to one rail pair (Table~\ref{tab:steane-noise}). An idle gap is an interval between a pair's preparation or reset and its readout in which no operation of nonzero duration touches either rail. Preparation and readout are otherwise ideal. Following Mehta et al.~\cite{mehta_bias-preserving_2025}, we assign half of the $\mathrm{CR}_z$ control's erasure probability before the gate.

\begin{table}[t]
  \centering
  \caption{Fault channels at each rail-pair location of the dual-rail Steane memory, with the erasure and residual probabilities $e_g$ and $\epsilon_g$ of Eqs.~\eqref{eq:hardware-loss-model} and~\eqref{eq:hardware-residual-model} and Table~\ref{tab:device-models}. $t_{\mathrm{idle}}$ is the length of an idle gap, and a routing SWAP's residual acts on the pair that ends on its first operand. The transmon rows of Table~\ref{tab:outer-code} use the Pauli-twirled thermal channel instead.}
  \label{tab:steane-noise}
  \footnotesize
  \setlength{\tabcolsep}{4pt}
  \begin{tabular}{@{}lll@{}}
    \toprule
    Location & Erasure & Residual \\
    \midrule
    Idle gap & $1-e^{-t_{\mathrm{idle}}/T_1}$ & $Z$, $(1-e^{-2\gamma t_{\mathrm{idle}}})/2$ \\
    $R_{\mathrm{SWAP}}$ & $e_{R_{\mathrm{SWAP}}}$ & $Z$, $\epsilon_{R_{\mathrm{SWAP}}}$ \\
    Routing SWAP & $e_{R_{\mathrm{SWAP}}}$ on each pair & $Z$, $\epsilon_{R_{\mathrm{SWAP}}}$ on one pair \\
    $\mathrm{CR}_z$ control & $e_{\mathrm{C}}/2$ before, $e_{\mathrm{C}}/2$ after & $Z$, $\epsilon_{\mathrm{C}}$ after \\
    $\mathrm{CR}_z$ target & $e_{\mathrm{T}}$ after & $Z$, $\epsilon_{\mathrm{T}}$ after \\
    Joint check & idle gap, $t_{\mathrm{idle}}=2.94\,\mu$s & idle $Z$, depolarizing 0.12\% \\
    Readout & none & none \\
    \bottomrule
  \end{tabular}
\end{table}

\textit{Leakage.} An erased pair is empty until a check or readout flags it~\cite{pavlovich_erasure_2026}. Until then, every operation that touches it is skipped as a unit, including a $\mathrm{CR}_z$ with it as either operand, the $R_Z$ corrections on the intact partner, and the pair's own residual faults. A control lost before its $\mathrm{CR}_z$ therefore leaves the target without a phase that depends on the lost control, so the target receives $Z$ with probability $1/2$, or $e_{\mathrm{C}}/4=10^{-3}$ per gate. A flag resets the pair to the completely mixed state, and a false flag resets an intact pair in the same way, because the hardware cannot tell the two apart. Ancilla pairs are also reset in every round, and data pairs are never otherwise reset. Stim has no leakage state, so each shot with a fault runs Stim's tableau simulator and skips the empty pair's operations.

\textit{Checks.} A check takes 2.94\,$\mu$s including readout. It flags an erased pair with probability $1-f_n$ and an intact one with probability $f_p$, with $f_n=3.7\%$ and $f_p=0.51\%$. A missed loss stays unreported until the pair's next check or readout, and a readout flags every unreported loss and never flags falsely. With an ideal flag at every loss instead, which resets each lost pair before it skips a gate, \sysname{} lowers $P_L$ by 25.7\% [18.5, 32.3] in the X memory and 25.4\% [17.1, 32.8] in the Z memory over the same twenty routes with 200,000 shots each.

\textit{Decoder.} A flag does not identify when its pair was lost, so for flag $k$ the decoder considers each fault location $\ell$ since the pair's previous flagged check, readout, or reset. With $e_\ell$ the erasure probability of $\ell$ and $a_\ell$ the number of unflagged checks between $\ell$ and the flag,
$w_\ell = e_\ell f_n^{a_\ell}\prod_{\ell'<\ell}(1-e_{\ell'})$
is the probability that the pair was first lost at $\ell$ and every later check missed it. For a loss at $\ell$, the possible errors are the fault mechanisms in Stim's detector error model after removing the pair's operations from $\ell$ to the flag and depolarizing it at the flag. Each such mechanism becomes a decoder column with prior
\begin{equation}
  q_k(\ell) = \frac{1}{2}\,
    \frac{w_\ell}
         {\sum_{\ell'} w_{\ell'} + c_k f_p/(1-f_n)},
  \label{eq:steane-prior}
\end{equation}
where $c_k$ is 1 for a check and 0 for a readout, so the $f_p$ term weighs a false flag. Contributions from several locations combine as \mbox{$p+q-2pq$}, and the reset's own mechanisms, which occur for every candidate loss location and also carry the false-flag weight, receive prior $1/2$. Without a flag, the decoder keeps each location's residual $Z$ prior and, at a check, two thirds of the depolarizing probability on each of its $X$ and $Z$ faults. BP+OSD runs seven serial product-sum iterations and order-2 combination-sweep OSD.

\section{AI Use}
The authors wrote the original software and manuscript. Claude (Anthropic) and ChatGPT (OpenAI) later assisted in revising both, and the authors reviewed and verified all AI-assisted changes.

\end{document}